\documentclass[a4paper,11pt]{article}
\usepackage{amsmath,amsthm,latexsym,amsfonts,graphicx,epsfig}
\usepackage[all]{xy}

\newcommand{\parang}[1]{\langle {#1} \rangle}
   \let\para\parang
\newcommand{\parto}[1]{\left(#1\right)}

\newcommand{\pos}{\Diamond}
\newcommand{\nec}{\Box}
\newcommand{\um}{\frac{1}{2}}

\newcommand{\IN}{{\mathbb N}}
\newcommand{\IQ}{{\mathbb Q}}
\newcommand{\IR}{{\mathbb R}} 
\newcommand{\logici}{\left\{ 0, \frac{1}{d-1}, \frac{2}{d-1}, \ldots,%
                     \frac{d-2}{d-1}, 1 \right\}}
\newcommand{\firstfunction}{\underline{f}_d^1}
\newcommand{\secondfunction}{\underline{f}_{d,\lambda}^2}

\newcommand{\mv}{\underline{m}_d}
\newcommand{\cA}{{\mathcal A}}
\newcommand{\IA}{{\mathbb A}}

\newtheorem{theorem}{Theorem}[section]
\newtheorem{lemma}{Lemma}[section]
\newtheorem{proposition}{Proposition}[section]

\newtheorem{definition}{Definition}[section]

\newenvironment{lquote}{\begin{list}{}{}\item[]}{\end{list}}

\begin{document}

\title{{\bf Fredkin Gates for Finite--valued \\ Reversible and Conservative Logics}}
\author{{\sc G. Cattaneo, A. Leporati, R. Leporini
\footnote{This work has been supported by MIUR$\backslash$COFIN project ``Formal Languages and
Automata: Theory and Application''}}
\\[0.3cm]
    Dipartimento di Informatica, Sistemistica e Comunicazione \\
    Universit\`a degli Studi di Milano -- Bicocca \\
    Via Bicocca degli Arcimboldi 8, 20126 Milano, Italy \\[0.3cm]
    e-mail: \{ cattang, leporati, leporini \}@disco.unimib.it }
\date{}
\maketitle
\begin{abstract}
The basic principles and results of \emph{Conservative Logic} introduced by Fredkin and Toffoli in
\cite{fredkin-toffoli} on the basis of a seminal paper of Landauer~\cite{landauer} are extended to
$d$--valued logics, with a special attention to three--valued logics. Different approaches to
$d$--valued logics are examined in order to determine some possible universal sets of logic primitives. In
particular, we consider the typical connectives of \L ukasiewicz and G\"odel logics, as well as Chang's MV--algebras. As
a result, some possible three--valued and $d$--valued universal gates are described which realize
a functionally complete set of fundamental connectives.
\end{abstract}

\section{Introduction}

The present paper is based on two different research areas which have been developed in the last
years: \emph{Conservative Logic\/} and \emph{Many--valued Logics}. Conservative logic is a model of
computation introduced by Fredkin and Toffoli in \cite{fredkin-toffoli} on the basis of the seminal
paper of Landauer~\cite{landauer} (see also~\cite{bennett-73}) to improve the efficiency and
performance of computing processes in terms of dissipated energy. The model is based on the
\emph{Fredkin gate}, a universal Boolean gate which is both conservative and reversible.

On the other hand, many--valued logics are extensions of classical two--valued (i.e.,
Boolean) logic which have a great diffusion due to their ability to manage incomplete
and/or uncertain knowledge. These two main subjects are briefly described in the next
sections.

In this paper we propose an extension of conservative logic in order to include the main features
of many--valued logics with a finite number of truth values. As a result we define some $d$--valued
universal gates which have the properties required by the conservative and many--valued paradigms.

\section{Reversibility, Conservativeness, and Conditional Control of Boolean Gates}

Computational models are usually based upon Boolean logic, and use some universal set of
primitive connectives such as, for example, $\{{\rm AND}, {\rm NOT}\}$.

From a general point of view, a ({\em classical deterministic}) {\em
$n$--inputs/$m$--outputs gate\/} (where $n,m$ are positive integers) is a
special--purpose {\em computer\/} schematized as a device able to compute (Boolean) {\em
logical functions\/}
$G:\{0,1\}^n\to\{0,1\}^m$.
Any $\Vec{x} = (x_1,x_2,\ldots,x_n)\in\{0,1\}^n$ (resp., $\Vec{y} =
(y_1,y_2,\ldots,y_m)\in\{0,1\}^m$) is called an {\em input\/} (resp., {\em output\/}) {\em
configuration\/}. For every $i\in\{1,2,\ldots,n\}$ (resp., $j\in\{1,2,\ldots,m\}$), called the {\em
input} (resp., {\em output\/}) {\it bit\/} of position $i$ (resp., $j$), the Boolean value
$x_i\in\{0,1\}$ (resp., $y_j\in\{0,1\}$) is said to be the {\em state\/} of bit $i$ (resp., $j$)
with respect to configuration $\Vec{x}$ (resp., $\Vec{y}$).
%
Finally, in the sequel we denote by $\lambda_f$ the generic configuration belonging to the range of
$G$.

\begin{figure}[ht]
   \setlength{\epsfxsize}{10cm}
   \begin{center}
      \mbox{\epsffile{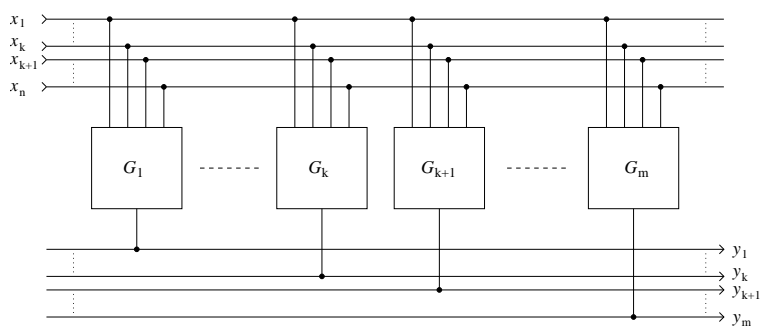}}
   \end{center}
   \caption{Standard parallel architecture of an $n$--inputs/$m$--outputs gate}
   \label{fig:pdp-1}
\end{figure}

The action of the multi--output map $G$ on an input configuration $\Vec{x}$ produces the
output configuration $G(\Vec{x})=(G_1(\Vec{x}),G_2(\Vec{x}),\ldots,G_m(\Vec{x}))$
determined by the component {\em logical truth functions\/} (single--output maps)
$G_j:\{0,1\}^n\to\{0,1\}$, for any $j=1,2,\ldots,m$, with a possible {\em parallel\/}
implementation drawn in Figure~\ref{fig:pdp-1}.

{\em Conservative logic\/} is a theoretical model of computation whose principal aim is to compute
with zero {\em internal power dissipation\/}. This goal is reached by basing the model upon {\em
reversible\/} and {\em conservative primitives\/}, which reflect physical principles such as the
reversibility of microscopic dynamical laws and the conservation of certain \emph{physical
quantities}, such as the energy of the physical system used to perform the computations.

{\bf Reversibility.\/}\quad Most of the times, computational models lack of \emph{reversibility};
that is, one cannot in general deduce the input values of a gate from its output values. For
example, knowing that the output of an AND gate is the logical value $0$ one cannot deduce the
input values that generated it. The original motivation for the study of reversibility in classical
computing came from the observation that heat dissipation is one of the major obstacles for
miniaturization of classical computers and the fact that the second law of thermodynamics implies
that irreversible state changes during computation must dissipate heat. ``Thus, in the more
abstract context of computing, the laws of ``conservation of information'' may play a role
analogous to those of conservation of energy and momentum in physics.'' \cite{toffoli}.

Lack of reversibility means that during the computation some information is lost.
As shown by R.~Landauer~\cite{landauer} (see also C.H.~Bennett \cite{bennett} which can be found
in~\cite{leff-rex}), a loss of information implies a loss of energy and therefore any computational
model based on irreversible primitives is necessarily {\em informationally dissipative\/}. This is
nowadays known (see \cite{be98b}) as:

\begin{quote}
{\bf Landauer's principle.} To erase a bit of classical information within a computer,
$1$ bit of entropy must be expelled into the computer's environment, typically in the
form of waste heat.
Thus logical irreversibility is associated with physical irreversibility and
requires a minimal heat generation, per machine cycle, typically of the order
of $kT$ for each irreversible operation.
\end{quote}
In practice the heat dissipation per bit processed by (irreversible) computers in use today is some
orders of magnitude greater than the theoretical lower bound $1kT\ln 2$ given by Landauer's
principle. However, if computer hardware continues to shrink in size as so far, then the only
feasible option to beat Landauer's lower bound seems to be reversible computation.\footnote{ In
modern computers heat dissipation is about $kT10^8$ per logical operation. The heat must be removed
by external means, for example, by constant cooling of all components by the thermal coupling of
the circuits to a heat reservoir, i.e., air.
 }

\begin{quote}
``In the customary approach, this transition [from the irreversibility of the given
computing process to the reversibility of the physical laws, NdA] occurs at a very low
level and is hidden --- so to speak --- in the ``physics'' of the individual gate; as a
consequence of this approach, the details of the work--to--heat conversion process are
put beyond the reach of the conceptual model of computation that is used.

On the other hand, it is possible to formulate a more general conceptual model of computation such that the gap between
the irreversibility of the desired behavior and the reversibility of a given underlying mechanism is bridged \emph{in
an explicit way} within the model itself.'' \cite[p.~3]{toffoli}.
\end{quote}

Let us make these considerations less informal by considering as a first example the logical
function $L$ of Table~\ref{tb:3-3-lan} computed by a three--inputs/three--outputs gate and
discussed by Landauer in \cite{landauer}.

\begin{table}[ht]
\begin{center}
\begin{tabular}{|c|c|c|c|c|c|c|}    
   \hline
   $x_1$ & $x_2$ & $x_3$ & $\to$ & $y_1$ & $y_2$ & $y_3$ \\
   \hline\hline
   \mbox{\rule[0cm]{0cm}{2.5ex}$0$} & $0$ & $0$ & & $0$ & $0$ & $0$ \\
   $0$ & $0$ & $1$ & & $1$ & $1$ & $0$ \\
   $0$ & $1$ & $0$ & & $0$ & $0$ & $0$ \\
   $0$ & $1$ & $1$ & & $1$ & $1$ & $0$ \\
   $1$ & $0$ & $0$ & & $0$ & $0$ & $0$ \\
   $1$ & $0$ & $1$ & & $1$ & $1$ & $0$ \\
   $1$ & $1$ & $0$ & & $0$ & $0$ & $1$ \\
   \mbox{\rule[0cm]{0cm}{1.5ex}$1$} & $1$ & $1$ & & $1$ & $1$ & $1$ \\
   \hline
\end{tabular}
\caption{The Landauer three--inputs/three--outputs gate}
\label{tb:3-3-lan}
\end{center}
\end{table}

Following Landauer ``There are eight possible initial states, and in thermal equilibrium they will
occur with equal probability. How much entropy reduction will occur in a machine cycle? States
$(111)$ and $(001)$ occur with a probability $1/8$ each; states $(110)$ and $(000)$ have a
probability of occurrence of $3/8$ each. The initial entropy was:
\begin{align*}
S_i(3)&=-k\sum_{\vec{x}} P(\vec{x})\log_e P(\vec{x}) \\
   &=-k\sum \frac{1}{8}\log_e \frac{1}{8}
    =3\,k\log_e 2
\end{align*}
The final entropy is
\begin{align*}
S_f(L)&=-k\sum_{\lambda_f} P_L(\lambda_f)\log_e P_L(\lambda_f) \\
   &=-k\Bigl(
  \frac{1}{8}\log_e \frac{1}{8}+\frac{1}{8}\log_e \frac{1}{8}+\frac{3}{8}\log_e \frac{3}{8}+\frac{3}{8}\log_e \frac{3}{8}
           \Bigr)
\end{align*}
The difference $S_i(3)-S_f(L)$ is $0.82\,k$. The minimum dissipation, if the initial
state has no useful information, is therefore $E_i(3)-E_f(L)=(S_i(3)-S_f(L)) \,T =
0.82\,k\,T$.''

More precisely, for any admissible output $\lambda_f=(y_1,y_2,y_3)\in Im(L)$ we can introduce the
set
$$
M_L(\lambda_f):=L^{-1}(\lambda_f)=\{(x_1,x_2,x_3)\in\{0,1\}^3:L(x_1,x_2,x_3)=\lambda_f\}
$$
whose cardinality $|M_L(\lambda_f)|$ expresses the {\em indistinguishability degree\/} of
the output $\lambda_f$, i.e., the total number of possible inputs which cannot be
distinguished by $L$ with respect to the output $\lambda_f$. Then the above probabilities
can be expressed as
\begin{equation*}
P_L(\lambda_f)=\frac{|M_L(\lambda_f)|}{|\{0,1\}^3|}=\frac{1}{8}\;|M_L(\lambda_f)|
\end{equation*}

We want now to extend these considerations in order to compare the dissipation
of {\em informational energy\/} in the case of devices whose number of output
lines is not necessarily equal to the number of input lines. To this end, let
us denote by ${\cal F}(\{0,1\},n,m)=(\{0,1\}^m)^{\{0,1\}^n}$ the collection of
all Boolean gates $G:\{0,1\}^n\to\{0,1\}^m$, and by
${\cal F}(\{0,1\},n,\IN)=\bigcup_{m\in\IN}{\cal F}(\{0,1\},n,m)$ the collection
of all Boolean gates with $n$ fixed and $m$ ranging in $\IN$.
For instance, ${\cal F}(\{0,1\},2,\IN)$ contains both the gate
${\rm AND}:\{0,1\}^2\to\{0,1\}$, associating to the Boolean pair $(x_1,x_2)$ the Boolean value
${\rm AND}(x_1,\,x_2)=x_1\cdot x_2$, and the two--inputs/four--outputs gate defined by Table~\ref{tb:4c-not}.

\begin{table}[ht]
\begin{center}
\begin{tabular}{|c|c|c|c|c|c|c|}
   \hline
   $x_1$ & $x_2$ & $\longmapsto$ & $y_1$ & $y_2$ & $y_3$ & $y_4$\\
   \hline\hline
   \mbox{\rule[0cm]{0cm}{2.5ex}$0$} & $0$ & & $0$ & $0$ & $0$ & $0$\\
   $0$ & $1$ & & $0$ & $1$ & $1$ & $0$\\   $1$ & $0$ & & $0$ & $1$ & $1$ & $1$\\
   \mbox{\rule[0cm]{0cm}{1.5ex}$1$} & $1$ & & $1$ & $0$ & $0$ & $1$\\
   \hline
\end{tabular}
\caption{Example of a two--inputs/four--outputs reversible gate}
\label{tb:4c-not}
\end{center}
\end{table}

In ${\cal F}(\{0,1\},n,\IN)$, owing to the assumption that in thermal equilibrium all possible
inputs $\vec{x}$  will occur with equal probability $P(\vec{x})$,  the {\em input information
entropy\/} is independent from the particular gate and equal to:
\begin{align*}
S_i(n):&=-k\sum_{\vec{x}} P(\vec{x})\log_e P(\vec{x}) \\
   &=-k\sum_{\vec{x}} \frac{1}{2^n}\log_e \frac{1}{2^n}
    =n\,k\,\log_e 2
\end{align*}
What depends on the gate $G\in {\cal F}(\{0,1\},n,\IN)$ is the set of $\lambda_f$--{\em indis\-tin\-gui\-sha\-ble input configurations\/}, where $\lambda_f\in Im(G)$ is any {\em admissible output configuration\/} of $G$:
\begin{align*}
M_G(\lambda_f):&=G^{-1}(\lambda_f)  \\
 &=\{(x_1,x_2,\ldots,x_n)\in\{0,1\}^n:G(x_1,x_2,\ldots,x_n)=\lambda_f\}
\end{align*}
Let us notice that the collection $\{M_G(\lambda_f):\lambda_f\in Im(G)\}$ of all such subsets
constitutes a partition of $\{0,1\}^n$. Borrowing some terminology from axiomatic quantum
mechanics, elements $\lambda_f$ from $Im(G)$ can be called {\em eigenvalues} (possible output
values) of $G$, $Im(G)$ is the {\em spectrum} of $G$, the set $M_G(\lambda_f)$ is the {\em
eigenspace} (set of possible inputs) associated to the eigenvalue $\lambda_f$, and the
characteristic function $\chi_{M_G(\lambda_f)}$ ($=1$ if $\vec{x}\in M_G(\lambda_f)$, and $0$
otherwise) is the spectral projection associated to the eigenspace. The collection of all spectral
projections of $G$, for $\lambda_f$ ranging on the spectrum of $G$, is a {\em spectral identity
resolution} of $G$:
\begin{gather*}
Id = \sum_{\lambda_f\in Im(G)} \chi_{M_G(\lambda_f)}
\\
G = \sum_{\lambda_f\in Im(G)} \lambda_f\,\chi_{M_G(\lambda_f)}
\end{gather*}

The {\em indistinguishability degree\/} of the admissible output configuration $\lambda_f\in Im(G)$ is defined as
$|M_G(\lambda_f)|$, and the {\em probability of occurrence\/} of $\lambda_f$ as:
\begin{equation*}
P_G(\lambda_f)=\frac{1}{2^n}\;|M_G(\lambda_f)|
\end{equation*}
with corresponding {\em output information entropy\/}:

\begin{subequations}
\label{eq:out-inf-ent}
\begin{align}
S_f(G):&=-k\sum_{\lambda_f\in Im(G)} P_G(\lambda_f)\log_e P_G(\lambda_f)  \\
    &=-\frac{k}{2^n} \sum_{\lambda_f\in Im(G)} |M_G(\lambda_f)|\cdot
\log_e|M_G(\lambda_f)|+ S_i(n)
\end{align}
\end{subequations}
Hence, the {\em information energy dissipation\/} of $G$ is:
\begin{align*}
\Delta E(G)&=\parto{S_i(n)-S_f(G)}\cdot T  \\
   &=\frac{kT}{2^n} \sum_{\lambda_f\in Im(G)} |M_G(\lambda_f)|\cdot \log_e|M_G(\lambda_f)|
\end{align*}
In particular, the information energy loss by the AND gate is $\Delta E({\rm
AND})=\frac{3kT}{4}\log_e 3 \approx 0.82\,k\,T$ whereas the gate of Table~\ref{tb:4c-not} (owing
to its reversibility) has no information energy dissipation.

From~\eqref{eq:out-inf-ent} it  follows immediately that the output information entropy
is bounded by:
\begin{equation*}
0\le S_f(G)\le S_i(n)
\end{equation*}
Of course, a generic gate $G:\{0,1\}^n\to\{0,1\}^m$ is {\em reversible\/} (one--to--one mapping)
iff $n=m$ and every element $\lambda_f$ of $\{0,1\}^n$ is an admissible output; in this case the
corresponding $|M_G(\lambda_f)|$ is equal to $1$ which leads to $S_i(n)-S_f(G)=0$, and thus also
$E_i(n)-E_f(G)=0$.
Precisely, the following Proposition holds.
\begin{proposition}
Let $G$ be any $n$--inputs Boolean gate. Then the information energy dissipation is bounded by
\begin{equation*}
0\le\Delta E(G)\le T\,S_i(n)=nkT\log_e 2
\end{equation*}
Moreover:
\begin{enumerate}
\item
$Im(G)$ is a singleton if and only if $\Delta E(G)=T\,S_i(n)$;
\item
the gate is reversible (one--to--one) if and only if $\Delta E(G)=0$.
\end{enumerate}
\end{proposition}

\noindent
Quoting Toffoli:
\begin{quote}
``\emph{Using invertible logic gates, it is ideally possible to build a sequential computer with
zero internal power dissipation}. The only source of power dissipation arises outside the circuit,
typically at the input/output interface, if the user chooses to connect input or output lines to
nonreversible digital circuitry. Even in this case, power dissipation is at most proportional to
the number of argument/result lines, rather than to the number of logic gates (as in ordinary
computers), and is thus independent of the ``complexity'' of the function being computed. This
constitutes the central result of the present paper.'' \cite[p.~32]{toffoli}
\end{quote}
Let us stress that in the case of an $n$--inputs/$n$--outputs gate realizing the logical function
$G:\{0,1\}^n\to\{0,1\}^n$ the reversibility condition corresponds to the fact that $G$ is a
permutation of the set $\{0,1\}^n$. For instance, a 2--inputs/2--outputs reversible gate computes a
permutation of the set $\{00,01,10,11\}$. Table~\ref{tb:22-perm} shows an example of a gate of this
kind.

\begin{table}[h]
\begin{center}
\begin{tabular}{|c|c|c|c|c|}
   \hline
   $x_1$ & $x_2$ & $\longmapsto$ & $y_1$ & $y_2$ \\
   \hline\hline
   \mbox{\rule[0cm]{0cm}{2.5ex}$0$} & $0$ & & $1$ & $1$ \\
   $0$ & $1$ & & $1$ & $0$ \\
   $1$ & $0$ & & $0$ & $1$ \\
   \mbox{\rule[0cm]{0cm}{1.5ex}$1$} & $1$ & & $0$ & $0$ \\
   \hline
\end{tabular}
\caption{Example of a two--inputs/two--outputs reversible gate, i.e., a
permutation of the set $\{0,1\}^2$}
\label{tb:22-perm}
\end{center}
\end{table}

{\bf Conservativeness.\/}\quad This condition is usually modelled by the property that each output
$(y_1,y_2,\ldots,y_n)$ of the gate is a permutation of the corresponding input
$(x_1,x_2,\ldots,x_n)$. We call this condition {\em strict conservativeness\/} of the gate.
Trivially a gate of this kind must necessarily have the same number of input and
output lines.  
In Table~\ref{tb:22-cons} an example of a (strictly) conservative $2$--inputs/$2$--outputs gate is
presented.

\begin{table}[ht]
\begin{center}
\begin{tabular}{|c|c|c|c|c|}
   \hline
   $x_1$ & $x_2$ & $\longmapsto$ & $y_1$ & $y_2$ \\
   \hline\hline
   \mbox{\rule[0cm]{0cm}{2.5ex}$0$} & $0$ & & $0$ & $0$ \\
   $0$ & $1$ & & $1$ & $0$ \\
   $1$ & $0$ & & $1$ & $0$ \\
   \mbox{\rule[0cm]{0cm}{1.5ex}$1$} & $1$ & & $1$ & $1$ \\
   \hline
\end{tabular}
\caption{Example of a two--inputs/two--outputs conservative gate, i.e., in each row the output pattern is a permutation
of the input pattern} \label{tb:22-cons}
\end{center}
\end{table}

\begin{quote}
``Some conservative (but not reversible) circuits using complementary signal streams were discussed
by von Neumann \cite{von-neumann} as early as 1952. More recently, Kinoshita and associates
\cite{kinoshita} worked out a classification of logic elements that ``conserve'' $0$'s and $1$'s;
their work, motivated by research in magnetic--bubble circuitry, mentions the possibility of more
energy--efficient computation, but has apparently little concern for reversibility.''
\cite[p.~33]{toffoli}
\end{quote}

The importance of conservativeness is further on stressed by Toffoli in \cite{toffoli}:
\begin{quote}
 ``In a conservative logic circuit, the number of $1$'s, which is conserved
in the operation of the circuit, is the sum of the number of $1$'s in different parts of the
circuit. Thus, this quantity is additive, and can be shown to play a formal role analogous to that
of energy in physical systems. [...]

In conclusion, conservative logic represents a substantial step in the development of a
model of computation that adequately reflects the basic laws of physics.''
\cite[p.~32]{toffoli}
\end{quote}
In general, in concrete devices the Boolean values $0$ and $1$ are realized by impulses
of energy $\varepsilon_0$ and $\varepsilon_1$ respectively, with $0<\varepsilon_0<\varepsilon_1$.
\begin{quote}
``In the classical realization the bit, which, for example could be imagined to be just a
mechanical switch, is a system which is designed to have two distinguishable states;
there should be a sufficiently large energy barrier between them that no spontaneous
transition, which would evidently be detrimental, can occur between the two
states.''~\cite{bo-ze}.
\end{quote}

\begin{figure}[ht]
   \setlength{\epsfxsize}{6cm}
   \begin{center}
      \mbox{\epsffile{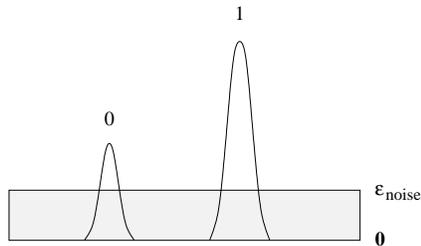}}
   \end{center}
   \caption{Realization of Boolean values 0 and 1 by impulses of energy
            $\varepsilon_0$ and $\varepsilon_1$, with
            $0 < \varepsilon_{\rm noise} < \varepsilon_0 < \varepsilon_1$}
   \label{fig:harm}
\end{figure}

In the case of a generic (non necessarily conservative) gate which computes a
logical function $G:\{0,1\}^n\to\{0,1\}^n$, a transition \break
$\Vec{x}=(x_1,x_2,\ldots,x_n)\to G(\Vec{x})=\Vec{y}=(y_1,y_2,\ldots,y_n)$ corresponding
to a row of the tabular definition of the Boolean function produces a variation of the
internal energy whose amount is
$$
\Delta U(\Vec{x},\Vec{y})=(\varepsilon_{y_1} + \varepsilon_{y_2} + \ldots +
\varepsilon_{y_n}) - (\varepsilon_{x_1} + \varepsilon_{x_2} + \ldots +
\varepsilon_{x_n})
$$
Therefore,  the total {\em internal energy dissipation\/} of $G$ is
\begin{equation*}
\Delta U(G)=\sum_{\Vec{x}\in\{0,1\}^n}\Delta U(\Vec{x},G(\Vec{x}))
\end{equation*}

Conservativeness of the gate $G$ trivially implies no internal energy dissipation ($\Delta
U(G)=0$).
\begin{quote}
``From the viewpoint of a physical implementation, where signals are encoded in some form of
energy, each constant input entails the supply of energy of predictable form, or \emph{work}, and
each garbage output entails the removal of energy of unpredictable form, or \emph{heat}. In this
context, a realization with fewer source and sink lines might point the way to a physical
implementation that dissipates less energy.'' \cite[p.~13] {toffoli}
\end{quote}
{\bf Conclusions on Reversibility and Conservativeness.}\quad
Up to now the loss of energy due to irreversibility and nonconservativeness of logical primitives
was irrelevant compared to the energy dissipated by an electronic device implementing logical
gates. But the problems rising from an extreme miniaturization in electronics have led to the
investigation of new ways of implementing circuits, borrowing the knowledge of quantum mechanics.
These new research areas introduce the possibility of reversible and conservative computations
based on reversible and conservative physical behavior, encouraging the definition of new
computational models.

Let us stress that there are gates which are reversible but non conservative (for instance, the
gate of Table~\ref{tb:22-perm} whose transition $00\to 11$ is non conservative) and gates which are
conservative and non reversible (for instance, the gate of Table~\ref{tb:22-cons} where both inputs
$01$ and $10$ are mapped into the same output $10$).

A simple example of reversible and conservative two--inputs/two--outputs  gate is the
realization of the {\em exchange\/} logical function ${\rm EXC}:\{0,1\}^2\to\{0,1\}^2$
whose tabular representation is given in Table~\ref{tb:exc}.
\begin{table}[ht]
\begin{center}
\begin{tabular}{|c|c|c|c|c|}
   \hline
   $x_1$ & $x_2$ & $\longmapsto$ & $y_1$ & $y_2$ \\
   \hline\hline
   \mbox{\rule[0cm]{0cm}{2.5ex}$0$} & $0$ & & $0$ & $0$ \\
   $0$ & $1$ & & $1$ & $0$ \\
   $1$ & $0$ & & $0$ & $1$ \\
   \mbox{\rule[0cm]{0cm}{1.5ex}$1$} & $1$ & & $1$ & $1$ \\
   \hline
\end{tabular}
\caption{The EXC two--inputs/two--outputs reversible and conservative gate}
\label{tb:exc}
\end{center}
\end{table}
In each row the output pair $(y_1,y_2)$ is a permutation of the corresponding input pair
$(x_1,x_2)$, and the map EXC is a (global) permutation of the set
$\{0,1\}^2=\{00,01,10,11\}$.

{\bf Conditional Control Gates.}\quad Let us consider the Boolean two--inputs/two--outputs
reversible non--conservative gate $G^{(CN)}:\{0,1\}^2\to\{0,1\}^2$ whose component maps are the
following:
\begin{align*}
&G^{(CN)}_1:\{0,1\}^2\to\{0,1\},\;\;G^{(CN)}_1(x_1,x_2):=x_1
\\
&G^{(CN)}_2:\{0,1\}^2\to\{0,1\},\;\; G^{(CN)}_2(x_1,x_2):=x_1\oplus x_2
\end{align*}
The corresponding truth table is given in Table~\ref{tb:c-not}.

\begin{table}[ht]
\begin{center}
\begin{tabular}{|c|c|c|c|c|}
   \hline
   $x_1$ & $x_2$ & $\longmapsto$ & $y_1$ & $y_2$ \\
   \hline\hline
   \mbox{\rule[0cm]{0cm}{2.5ex}$0$} & $0$ & & $0$ & $0$ \\
   $0$ & $1$ & & $0$ & $1$ \\
   $1$ & $0$ & & $1$ & $1$ \\
   \mbox{\rule[0cm]{0cm}{1.5ex}$1$} & $1$ & & $1$ & $0$ \\
   \hline
\end{tabular}
\caption{The Controlled--NOT reversible non--conservative gate}
\label{tb:c-not}
\end{center}
\end{table}

We can describe the behavior of this gate by considering the information $x_1$ as a
\emph{control input\/} which is left unchanged but which determines the action of a
prescribed operation on the {\em target input\/} $x_2$, transforming it into the output $y_2$.
To be precise, if the control input is 1 then the value of the target line is negated (i.e., the
gate NOT acts on $x_2$ when $x_1=1$), otherwise it is left unchanged (i.e., the gate Id acts on
$x_2$ when $x_1=0$). Formally, this is realized by a direct connection of the first input line with
the first output line, whereas the action on the input of the second line is described by two maps
$\delta^{(CN)}_{x_1}:\{0,1\}\to\{0,1\}$, where $\delta^{(CN)}_{x_1}=G^{(CN)}_2(x_1,\;\cdot\;)$ for
$x_1\in\{1,0\}$. Precisely,
\begin{equation*}
\delta^{(CN)}_{0}:=G^{(CN)}_2(0,\;\cdot\;)=\text{Id}\quad\text{and}\quad
\delta^{(CN)}_{1}:=G^{(CN)}_2(1,\;\cdot\;)=\text{NOT}
\end{equation*}
The input value of the control unit $x_1$ selects the map $\delta^{(CN)}_{x_1}$ (either the
identity or the NOT map) which acts on the input value $x_2$ of the second line. For this reason
this gate is called the \emph{Controlled--}NOT (usually abbreviated with CNOT) gate.

\begin{figure}[ht]
\begin{center}
   \includegraphics{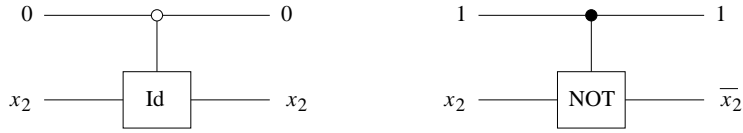}
\end{center}
\caption{The conditional action of the Controlled--NOT gate}
\label{fig:CNOT}
\end{figure}

\begin{figure}[ht]
\begin{center}
   \includegraphics[width=11.5cm]{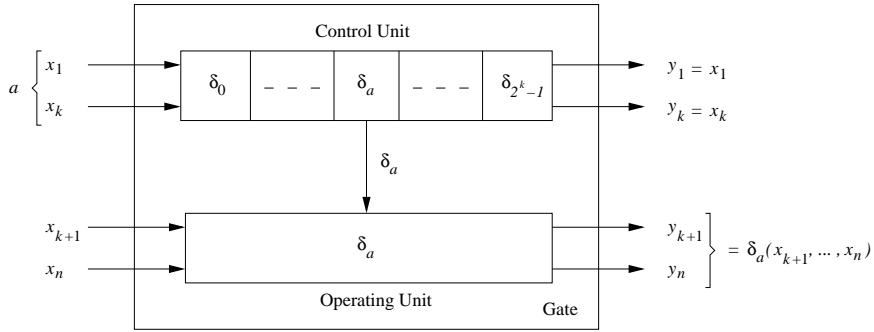}
\end{center}
\caption{Ideal realization of a generic Conditional Control gate: the gate is
         divided into a \emph{control unit} and an \emph{operating unit}.
         The input values of the control unit are left unchanged and select
         a prescribed function to be applied to the input values of the operating
         unit}
\label{fig:CQD-scheme}
\end{figure}

From the general viewpoint, the {\em Conditional Control\/} method applies to the cases in which
the $n$--inputs/$n$--outputs gate can be divided into two parts: a \emph{control unit} and an
\emph{target\/} (also {\em operating\/}) {\em unit\/} (see Figure~\ref{fig:CQD-scheme}).

The control unit has in general $k$ input and $k$ output lines, while the target unit has $(n-k)$
input and $(n-k)$ output lines. Thus any configuration ${x_1,\ldots,x_k,x_{k+1},\ldots,x_n}$ can be
split into two parts: the {\em control configuration\/} ${x_1,\ldots,x_k}$ and the {\em operating}
(sometimes also called {\em target}) {\em configuration\/} ${x_{k+1},\ldots,x_n}$. Any of the $2^k$
possible control configurations ${x_1,\ldots,x_k}$ is labelled by the integer number
$a=\sum_{t=1}^k x_t\,2^{t-1}$.
Moreover, $2^k$ functions $\delta_0, \delta_1,\ldots, \delta_{2^{k}-1}$ of the kind $\{0,1\}^{n-k}\to\{0,1\}^{n-k}$ are
stored in the memory of the control unit, the function $\delta_a$ being bijectively associated to the configuration
labelled by the integer number $a\in\{0,\ldots,2^{k}-1\}$.

When a configuration ${x_1, \ldots, x_k}$ (labelled by $a$) is fed as input to the control lines two things happen:
\begin{enumerate}
\item
the control configuration $x_1,\ldots,x_k$ is returned unchanged into the output lines of the control unit;
\item
the function $\delta_a$ bijectively associated to the control configuration is selected and applied to the input
configuration ${x_{k+1},\ldots,x_n}$ of the operating unit, producing the output configuration
$\delta_a(x_{k+1},\ldots,x_n)$.
\end{enumerate}
We can look at a controlled gate as a finite automaton. The original space $\{0,1\}^n$ on which a
controlled gate $G$ acts can be split in the set $\cA:=\{0,1\}^k$, called the {\em alphabet} of the
gate, and the set $Q:=\{0,1\}^{n-k}$, called the {\em phase} space of the gate; elements of $\cA$
are {\em symbols} of the alphabet and elements of $Q$ are {\em states} of the gate. Hence, the gate
can be represented as a mapping $G:\cA\times Q\to \cA\times Q$, associating with any symbol--state
pair $(\vec a,\vec s)$ a new symbol-state pair $G(\vec a,\vec s):=(\vec a,\delta_a(\vec s))$.
Therefore, if we put the gate in cascade with the trivial {\em decoder} (according to
\cite{toffoli}) $\pi_Q:\cA\times Q\to Q$ associating with any pair $(a,s)$ the single state
$\pi_Q(a,s):=s$ one obtains a deterministic finite {\em automaton} $\IA^G=\para{\cA,Q,\delta}$ with
(finite) alphabet $\cA$, set of states $Q$, and {\em next state} (also {\em transition}) {\em
function} $\delta:=(\pi_Q\circ G):\cA\times Q\to Q$ associating with any letter--state pair $(\vec
a,\vec s)$ the ``next'' state ${\vec s}\,' = \delta(\vec a,\vec s) := \pi_Q(G(\vec a,\vec s)) =
\delta_a(\vec s)$.

\begin{figure}[ht]
\begin{center}
   \includegraphics[width=8cm]{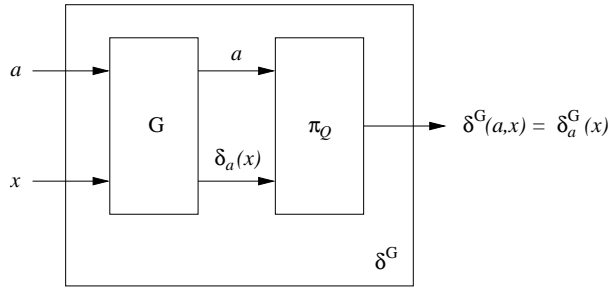}
\end{center}
\caption{Automaton generated by a controlled gate}
\end{figure}

This automaton can be equivalently described by the pair \break
$\para{Q,\{\delta_0,\delta_1,\ldots,\delta_{2^k -1}\}}$ consisting of the phase space
$Q=\{0,1\}^{n-k}$ and the collection of $2^k$ transformations of the phase space $\delta_a:Q\to Q$,
for $a$ running in $\{0,1,\ldots, 2^k-1\}$.

Vice versa, any (finite) automaton $\IA=\para{\cA,Q,\delta}$ consisting of the (finite) alphabet
$\cA$, the (finite) phase space $Q$, and the next state function $\delta:\cA\times Q\to Q$ can be
equivalently described by the pair $\para{Q,\{\delta_0,\delta_1,\ldots,\delta_{|\cA|-1}\}}$ based
on the phase space $Q$ and the (finite) collection of phase space transformations $\delta_a:Q\to Q$
(for $a\in\{0,1,\ldots,|\cA|-1$) associating with any state $\vec s$ the next state $\vec s\,' =
\delta_a(\vec s) :=\delta(\vec a,\vec s)$. This automaton generates a controlled gate
$G^\delta:\cA\times Q\to\cA\times Q$ associating with the symbol--state input pair $(\vec a,\vec
s)$ the symbol--state output pair $G^\delta(\vec a,\vec s):=(\vec a,\delta(\vec a,\vec s))$.
Trivially, if $|A|=2^k$ and $|Q|=2^h$ by a suitable binary representations of each symbol $\vec a$
and each state $\vec s$ this conditional control gate is realized by a mapping $G^\delta:
\{0,1\}^n\to\{0,1\}^n$, with $n=h+k$.

\begin{figure}[ht]
\begin{center}
   \includegraphics[width=7cm]{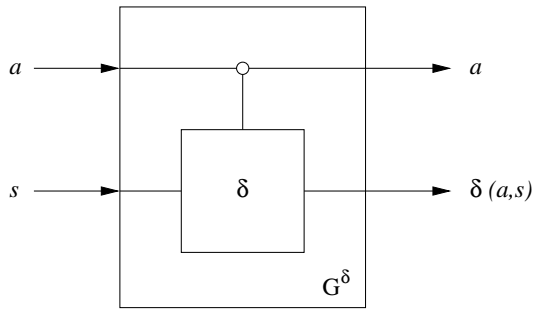}
\end{center}
\caption{Controlled gate generated by an automaton}
\end{figure}

In conclusion, the class of Boolean conditional control gates is categorically equivalent to the
class of (deterministic, finite) automata in which both the alphabet and the phase space have a
power of 2 cardinality.

\begin{figure}[ht]
\begin{center}
   \includegraphics[width=7cm]{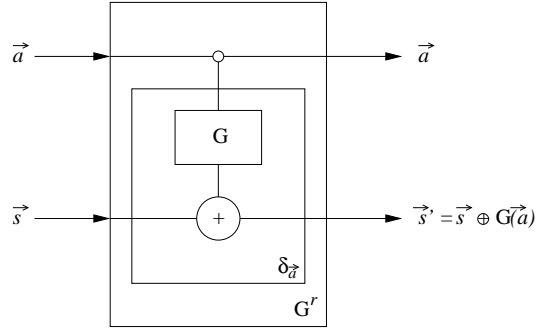}
\end{center}
\caption{Reversible gate generated by a nonreversible one}
\label{fg:non-to-rev}
\end{figure}

{\bf The reversible and conditional controlled gate generated by a nonreversible gate}.\quad
 If a Boolean gate $G:\{0,1\}^n \to \{0,1\}^m$ is not reversible, it is always possible to construct a corresponding
 reversible gate $G^r:\{0,1\}^{m+n} \to \{0,1\}^{m+n}$ associating to the input pair
 $(\vec a,\vec s)\in\{0,1\}^n \times \{0,1\}^m$ the output pair
 $(\vec a,\vec s\oplus G(\vec a))\in\{0,1\}^n \times \{0,1\}^m$. This gate is reversible (and generally non
 conservative). Moreover it is a controlled gate, that is a finite automaton with respect to the alphabet $\cA=\{0,1\}^n$,
 the phase space $Q=\{0,1\}^m$, and the set of next state functions $\delta_{\vec a}$ (for $\vec a\in\{0,1\}^n$)
 associating to any state $\vec s\in\{0,1\}^m$ the next state ${\vec{s} }\,^\prime=\delta_{\vec a}(\vec s):=\vec s\oplus G(\vec
 a)\in\{0,1\}^m$ (see Figure~\ref{fg:non-to-rev}).

{\bf How to transform a reversible and non conservative Boolean gate into
          a reversible and conservative one}.\quad
If $G^r$ is a non conservative reversible gate, we can extend it to a conservative gate $G^{rc}$ by
adding some new input and output lines, and maintaining the original reversibility. Let $\{\vec
x\}_1 = \sum_{i=1}^{n+m} x_i$ be the number of ones contained into the input $\vec x$; analogously,
let $\{G^r(\vec x)\}_1 = \sum_{i=1}^{n+m} G^r_i(\vec x)$ be the number of ones contained into the
corresponding output $G^r(\vec x)$. We denote with $E(\vec x)$ the quantity $\{G^r(\vec x)\}_1 -
\{\vec x\}_1$. Clearly $E(\vec x)$ is an integer number from the interval $[-(n+m), n+m]$. It is
immediately seen that if $G^r$ would be conservative then it would hold $E(\vec x) = 0$ for every
$\vec x \in \{0,1\}^{n+m}$. On the other hand, since we have supposed that $G^r$ is a non
conservative gate, there exists an $\vec x \in \{0,1\}^{n+m}$ such that $E(\vec x) \neq 0$.

For the moment, let us suppose that $E(\vec x) > 0$. Then there exists an $\vec x' \in
\{0,1\}^{n+m}$ such that $E(\vec x') < 0$. In fact we can express the quantity $\sum_{\vec x \in
\{0,1\}^{n+m}}E(\vec x)$ as follows:
\begin{align}
   \sum_{\vec x \in \{0,1\}^{n+m}} E(\vec x)
   &= \sum_{\vec x \in \{0,1\}^{n+m}} \Big(\{G^r(\vec x)\}_1 -
      \{\vec x\}_1\Big) \notag \\
   &= \sum_{\vec x \in \{0,1\}^{n+m}}\{G^r(\vec x)\}_1 -
      \sum_{\vec x \in \{0,1\}^{n+m}}\{x\}_1
      \label{eq:sums}
\end{align}
Since $G^r$ is reversible, it is a permutation over the set $\{0,1\}^{n+m}$. This means that the
two sums in \eqref{eq:sums} are over the same elements, and thus:
\begin{equation*}
   \sum_{\vec x \in \{0,1\}^{n+m}} E(\vec x) = 0
\end{equation*}
As a consequence, if $E(\vec x) > 0$ there must exist an $\vec x' \in \{0,1\}^{n+m}$ such that
$E(\vec x') < 0$. In a completely analogous way we can show that if $E(\vec x) < 0$ then there
exists an $\vec x' \in \{0,1\}^{n+m}$ such that $E(\vec x') > 0$.

For the considerations above, if we define $\ell = -\min_{\vec x} E(\vec x)$ and $h = \max_{\vec x}
E(\vec x)$, and the gate $G^r$ is non conservative, then $\ell$ and $h$ are positive integers. For
any $\vec x \in \{0,1\}^{n+m}$ such that $E(\vec x) < 0$, let $E_\ell(\vec x)$ be the string
$\underbrace{1,\ldots,1}_{- E(\vec x)}, 0,\ldots,0$ of length $\ell$ (if $\ell = 0$ we obtain the
empty string); analogously, whenever $E(\vec x) > 0$ we define $E_h^c(\vec x)$ as the string
$\underbrace{0,\ldots,0}_{E(\vec x)}, 1, \ldots,1$ of length $h$.

To extend $G^r$ to a reversible and conservative gate $G^{rc}$ we can use $\ell$ ancillae lines
(that we briefly indicate with $\vec y$) to provide $- E(\vec x)$ ones whenever $E(\vec x) < 0$,
and $h$ ancillae lines (that we indicate with $\vec z$) to remove $E(\vec x)$ ones whenever $E(\vec
x) > 0$. More precisely, we define $G^{rc}:\{0,1\}^{n+m+\ell+h} \to \{0,1\}^{n+m+\ell+h}$ as
follows:
\begin{align*}
   & \forall \, \vec x \in \{0,1\}^{n+m}, \,
     \forall \, \vec y \in \{0,1\}^\ell, \,
     \forall \, \vec z \in \{0,1\}^h \\
   & G^{rc} (\vec x,\vec y,\vec z) :=
     \begin{cases}
        (G^r(\vec x), E_\ell(\vec x), \vec 1_h) &
           \text{if $E(\vec x) < 0$, $\vec y = \vec 0$ and $\vec z = \vec 1$
                 \hspace{0.95cm} i)} \\
        (\vec k, \vec 0_\ell,\vec 1_h) &
           \text{if $G^r(\vec k) = \vec x$, $E(\vec k) < 0$,} \\
   & \hspace{0.4cm}\text{$\vec y = E_\ell(\vec k)$ and $\vec z = \vec 1$
           \hspace{1.9cm} ii)} \\
        (G^r(\vec x), \vec 0_\ell, \vec 1_h) &
           \text{if $E(\vec x) = 0$, $\vec y = \vec 0$ and $\vec z = \vec 1$
           \hspace{0.9cm} iii)} \\
        (G^r(\vec x), \vec 0_\ell, E_h^c(\vec x)) &
           \text{if $E(\vec x) > 0$, $\vec y = \vec 0$ and $\vec z = \vec 1$
           \hspace{0.9cm} iv)} \\
        (\vec k, \vec 0_\ell, \vec 1_h) &
           \text{if $G^r(\vec k) = \vec x$, $E(\vec k) > 0$,} \\
   & \hspace{0.4cm}\text{$\vec y = \vec 0$ and $\vec z = E_h^c(\vec k)$
           \hspace{1.8cm} v)} \\
        (\vec x,\vec y,\vec z) & \text{otherwise \hspace{4.1cm} vi)}
     \end{cases}
\end{align*}

A direct inspection of $G^{rc}$ shows that the map $G^r$ is obtained in the first $n+m$ output
lines when the ancillae lines $\vec y$ and $\vec z$ are fixed respectively with the input values
$\vec 0$ and $\vec 1$. Notice that the rules ii) and v) are designed in order to provide the
inverses of the tuples produced by rules i) and iv), respectively. On the other hand, the tuples
produced by rule iii) can be inverted by computing the inverse of the first $n+m$ components
through the inverse of the map $G^r$. Finally, rule vi) makes the gate behave as the identity when
none of the previous rules are satisfied: as a consequence, the corresponding tuples can be
trivially inverted. Summarizing, the inverse of $G^{rc}$ is obtained by substituting rule iii) in
the analytic expression of $G^{rc}$ with the following:
\begin{equation*}
   (\vec k, \vec 0_\ell, \vec 1_h) \qquad \text{if $G^r(\vec k) = \vec x$,
      $E(\vec k) = 0$, $\vec y = \vec 0$ and $\vec z = \vec 1$}
\end{equation*}
{\bf Reconstruction of the original gate from the reversible and conservative induced gate}.\quad
 Following Toffoli (\cite{toffoli}), the original arbitrary Boolean gate $G$
can be recovered by means of the just constructed reversible and conservative gate $G^{rc}$ in the
following way.

\begin{quote}
In more general mathematical parlance, a {\em realization} of a function $G$ consists in a new
function $G^{rc}$ together with two mappings $\mu$ and $\pi_Q$ (respectively, the {\em encoder} and
the {\em decoder}) such that $G=\pi_Q\circ G^{rc}\circ \mu$. In this context, our plan is to obtain
a realization $\pi_Q\circ G^{rc}\circ \mu$ of $G$ such that $G^{rc}$ is invertible [i.e.,
reversible] and conservative, and the mappings $\mu$ and $\pi_Q$ are essentially independent of $G$
and contain as little ``computing power'' as possible.

More precisely, though the form of $\mu$ and $\pi_Q$ must obviously reflect the number of input and
output components of $G$, and thus the format of $G$'s truth table, we want them to be otherwise
independent of the particular contents of such truth table as $G$ is made to range over the set of
all combinatorial functions.
\end{quote}

In the present case, the encoder is realized by the mapping $\mu : \{0,1\}^n \to
\{0,1\}^{n+m+\ell+h}$ associating to the input $\vec a \in \{0,1\}^n$ the output $4$--tuple
$\mu(\vec a) := (\vec a, \vec 0_m, \vec 0_\ell, \vec 1_h) \in \{0,1\}^n \times \{0,1\}^m \times
\{0,1\}^\ell \times \{0,1\}^h$ (independent of the particular form of $G$). The decoder is realized
by the projection mapping $\pi_Q : \mathcal{A} \times Q \times \{0,1\}^\ell \times \{0,1\}^h \to
Q$. Trivially, for any $\vec a \in \{0,1\}^n$ one gets $(\pi_Q \circ G^{rc}\circ \mu)(\vec a) =
(\pi_Q \circ G^{rc})(\vec a, \vec 0_m,
\vec 0_\ell, \vec 1_h) = \pi_Q(\vec a, G(\vec a), \vec y, \vec z) = G(\vec a)$.

\begin{figure}[ht]
\begin{center}
   \includegraphics[width=8cm]{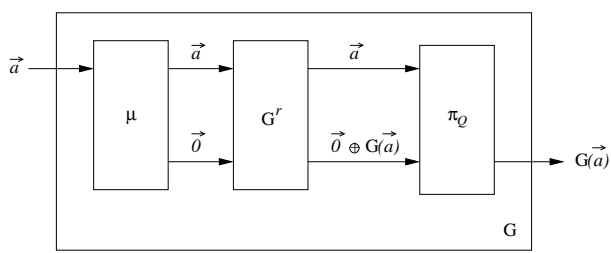}
\end{center}
\caption{Original nonreversible gate obtained by its reversible extension}
\end{figure}

\begin{figure}[ht]
\begin{center}
   \includegraphics{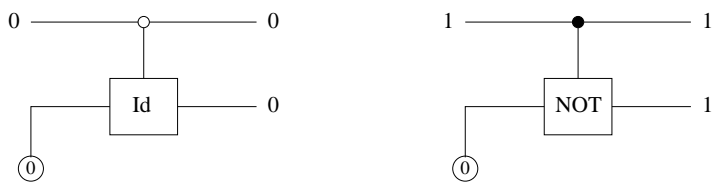}
\end{center}
\caption{Realization of the FAN--OUT function with the Controlled--NOT gate} \label{fig:CNOT-FANOUT}
\end{figure}

{\bf The FAN--OUT gate as a cloning procedure induced by the Controlled--Not gate.}\quad A very important connective in
reversible computing is $\text{FAN--OUT}: L \to L^2$, defined by the law $\text{FAN--OUT}(x) = (x, x)$. In other words,
the FAN--OUT function simply clones the input value. When dealing with classical circuits, the FAN--OUT function is
implemented by sticking two output wires to an existing input wire. The Controlled--NOT gate (see Table~\ref{tb:c-not})
provides a possible realization of the FAN--OUT function by a two--inputs/two--outputs reversible gate. Indeed, if the
operating line is fixed with the input value ${x_2=0}$, then the control input is cloned realizing in this way a
classical FAN--OUT (see Figure \ref{fig:CNOT-FANOUT}).

\section{The Conservative and Reversible Fredkin Gate}
{\bf (F1)}\quad
One of the paradigmatic conservative and reversible primitive is the {\it Fredkin gate}, a three--inputs/three--outputs gate that computes the following function
${\rm FG}:{\{0,1\}^3\rightarrow\{0,1\}^3}$:
\begin{gather*}
   y_1 = x_1 \\
   y_2 = (x_1 \land x_2) \lor (\lnot x_1 \land x_3) \\
   y_3 = (\lnot x_1\land x_2) \lor (x_1\land x_3)
\end{gather*}
In tabular notation it is presented in Table~\ref{tb:fredkin}.

\begin{table}[ht]
\begin{center}
\begin{tabular}{|c|c|c|c|c|c|c|}
   \hline
   $x_1$ & $x_2$ & $x_3$ & $\longmapsto$ & $y_1$ & $y_2$ & $y_3$ \\
   \hline\hline
   \mbox{\rule[0cm]{0cm}{2.5ex}$0$} & $0$ & $0$ & & $0$ & $0$ & $0$\\
   $0$ & $0$ & $1$ & & $0$ & $1$ & $0$ \\
   $0$ & $1$ & $0$ & & $0$ & $0$ & $1$ \\
   $0$ & $1$ & $1$ & & $0$ & $1$ & $1$ \\
   $1$ & $0$ & $0$ & & $1$ & $0$ & $0$ \\
   $1$ & $0$ & $1$ & & $1$ & $0$ & $1$ \\
   $1$ & $1$ & $0$ & & $1$ & $1$ & $0$ \\
   \mbox{\rule[0cm]{0cm}{1.5ex}$1$} & $1$ & $1$ & & $1$ & $1$ & $1$\\
   \hline
\end{tabular}
\caption{The Fredkin reversible and conservative gate}
\label{tb:fredkin}
\end{center}
\end{table}

{\bf (F2)}\quad The Fredkin gate is self--reversible, i.e., the inverse of
function FG is FG itself. This is a particularly desirable feature for the
construction of the quantum version of a reversible circuit, since the part of
the circuit which ``undoes'' the computation (in order to disentangle
input/output lines and the so--called ancillae lines) is thus completely
symmetrical to the part which computes the output value.

Note that self--reversibility implies the reversibility property.
%
%
The converse is not generally true: if $f : L^m \to L^m$ is reversible then it is a permutation of $L^m$ and, as it is well known, in general the composition of a permutation with itself does not give the identity as a result.
In particular, it is immediate to see that only those permutations which are
expressible as the composition of disjoint cycles of length $2$ (and fixed
points) are self--reversible.

{\bf (F3)}\quad
Looking at Table~\ref{tb:fredkin}, it follows immediately that the Fredkin gate is conservative.
This property allows for the realization of the Fredkin gate in the framework
of ``billiard ball'' computing, and led to the following observation concerning
the physical meaning of conservativeness:
\begin{quote}
   \label{cit:fredkin-toffoli}
   ``In conservative logic, all signal processing is ultimately reduced to
   \emph{conditional routing} of signals.
   Roughly speaking, signals are treated as unalterable objects that can be
   moved around in the course of a computation but never created or
   destroyed.'' (\cite{fredkin-toffoli}, page 227).
\end{quote}
The billiard ball model, developed by Fredkin and Toffoli in \cite{fredkin-toffoli}, is
an excellent example of a toy scientific model of no immediate practical application but
of large scientific impact.
Balls of radius $1$ travel on a unit grid in two directions. The direction of their movements can
be changed either by an elastic collision, or by a reflection at a ``mirror''. Using this model it
is possible to implement a \emph{switch gate}; the Toffoli gate can then be implemented with four
of them.

\begin{figure}[ht]
   \begin{center}
      \mbox{\epsffile{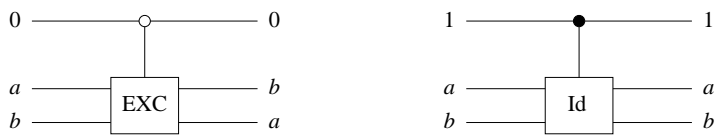}}
   \end{center}
   \caption{The Fredkin gate as a conditional switch}
   \label{fig:conditional-switch}
\end{figure}

{\bf (F4)}\quad If the first of the inputs is set to $0$ then the Fredkin gate exchanges
the second input with the third one, whereas if the first input is set to $1$ it returns
all the inputs unchanged, as it is shown in Figure~\ref{fig:conditional-switch}.

Therefore, the Fredkin gate is a Conditional Control gate with $x_1$ as control input,
$\delta_0=$EXC and $\delta_1=$Id.

{\bf (F5)}\quad From the Fredkin gate we can obtain some classical unary and binary
connectives by setting respectively two and one of the input lines to a constant value
(that is, either $0$ or $1$). For example,
\begin{itemize}
\item
by fixing $x_3 = 0$ in the input, the second output becomes $y_2 = (x_1 \land x_2) \lor
(\lnot x_1 \land 0) = x_1 \land x_2$, i.e., $y_2$ gives the logical AND between $x_1$
and $x_2$. In this case the outputs $y_1$ and $y_3$ are called {\it garbage};
\item
by fixing $x_2=1$ and $x_3=0$ the input $x_1$ is negated in the output
$y_3=\neg x_1$, with garbage $y_1$ and $y_2$. In this way we realize the NOT
connective.
\end{itemize}

{\bf (F6)}\quad Differently from the realization of the FAN--OUT gate by the reversible
nonconservative Controlled--NOT gate, it is easy to see that it is impossible to realize
the FAN--OUT function by a conservative two--inputs/two--outputs gate. Such a realization
requires at least three input lines and three output lines, even when working with
Boolean logic. The Fredkin gate supplies one of these possible realizations:

\begin{itemize}
\item
if we fix $x_2=1$ and $x_3=0$ then the first input is cloned in the first and second
outputs, i.e., we obtain the FAN--OUT function, with the output $y_3$ as garbage.
\end{itemize}

Summarizing, the Fredkin gate has the following properties:
\begin{lquote}
\begin{itemize}
   \item[F-1~)] it is a three--inputs/three--outputs gate, where each
                input/output assumes values in $\{ 0, 1\}$;
   \item[F-2~)] it is {\it reversible}, that is a bijective mapping from
                $\{0,1\}^3$ onto $\{0,1\}^3$;
   \item[F-2')] it is {\it self--reversible}, that is ${\rm FG}^2 = {\rm FG}
                \circ {\rm FG} = \text{Id}$ (the identity function on $\{0,1\}^3$);
   \item[F-3~)] it is {\it conservative}, in the sense that the number of $0$
                and $1$ in the input is the same as in the output;
   \item[F-4~)] it is a {\em controlled gate\/}, that is $x_1$ is a
                {\em control input\/} which is left unchanged but which determines a
                transformation of the {\em target input\/} $(x_2,x_3)$
                into the output $(y_2,y_3)$ by the gate EXC if $x_1=0$ and by the
                identity gate if $x_1=1$;
   \item[F-5~)] it is a {\it universal primitive}, that is, from the
                configurations of the gate we can obtain the classical logical
                connectives AND, OR, IMPLICATION, NOT which constitute a
                ``functionally complete'' set of connectives for the Boolean logic,
                that is a set of primitive truth functions with which
                all the possible truth functions (i.e., all the functions $\{0, 1\}^n\to\{0, 1\}$
                for $n$ ranging in $\IN$) can be realized;
   \item[F-6~)] it realizes the FAN--OUT connective, which plays
                a central role in reversible computations since it clones a
                given input signal.
\end{itemize}
\end{lquote}

Our aim is to extend this computational Boolean framework based on the Fredkin gate to include the
main features of many--valued logics, when a finite number of truth values are involved. In the
next section we give a brief summary of the main aspects of this subject.

\section{Many--valued Logics \label{section:MVLogic}}

The simplest extension of classical two--valued logic consists in the
introduction of a third  ``intermediate'', or ``neutral'' or ``indeterminate''
value.
\L ukasie\-wicz  developed this idea in \cite{lukasiewicz}. In such paper he introduced a third
truth value to take into account propositions which are neither true nor false, defining in this
way a three--valued logic. This logic was then extended to deal with $d$ truth values as well as
with an infinite number of truth values, in particular the $\aleph_0$ and $\aleph_1$ cardinalities.

Let us begin with a brief exposition of the main features of the many--valued
logics of \L ukasiewicz; the definition and the properties of the operators
are the same for the finite and the infinite--valued cases, unless otherwise
stated.

Technically speaking, truth values of a logical system are defined just as
syntactic labels, with no numerical meaning.
In a subsequent step, it is possible to give an interpretation of the logical
system in terms of an algebraic structure; only during such a process, the
truth values are associated with elements of the structure, which can be more
abstract mathematical objects than real or integer numbers.
Indeed, all the notions here exposed can be restated in such a formal way;
however, for our purposes it will be convenient to deal with the following
sets of truth values, treated as numerical sets equipped with the standard
total order relation induced by $\IR$:
\begin{itemize}
   \item $L_d = \logici$, with $d \ge 2$, for $d$--valued logics;
   \item $L_{\aleph_0} = [0,1] \cap \IQ$, that is the set of rational in the
         interval $[0,1]$, for infinite--valued logics with $\aleph_0$ truth
         values;
   \item $L_{\aleph_1} = [0,1]$, that is the set of real values in the interval
         $[0,1]$, for infinite--valued logics with $\aleph_1$ truth values.
\end{itemize}
The numbers of $L_\alpha,\;\alpha\in\{d,\aleph_0,\aleph_1\}$ are interpreted, after \L
ukasiewicz, as the possible truth values which the logical sentences can be assigned to.
As usually done in literature, the values $1$ and $0$ denote respectively truth and
falseness, whereas all the other values are used to indicate different degrees of
indefiniteness. With the introduction of the new truth values, the propositional
connectives of Boolean logic must be redefined. Accordingly, many--valued logics
represent strong generalizations of bivalent (i.e., classical) logic.
\subsection{\L ukasiewicz approach}
The {\bf \L ukasiewicz\/} system on the totally ordered numerical set of truth values
$L_\alpha$, with $\alpha\in\{d,\aleph_0,\aleph_1\}$, considers as primitive the {\em
implication\/} ($\to_L$) connective, which is defined by the following equation:

\begin{align*}
   x\to_L y :&=\min\{1,1-x+y\}& \text{(\L ukasiewicz~implication)} \\
             &= \begin{cases}
                   1 - x + y & \text{if \ $y < x$} \\
                   1         & \text{otherwise}
                \end{cases} \\
\end{align*}
In the system $\para{L_\alpha,\to_L\,}$ a {\em negation\/} ($\lnot$) connective is derived according to the rule:
\begin{align*}
   \lnot x :&= x \to_L 0     & \text{(diametrical~negation)} \\
            &= 1 - x
\end{align*}
Using these two connectives \L ukasiewicz defines some other derived ones as:
\begin{align*}
   x \lor y &:= (x \to_L y) \to_L y & \text{(\L ukasiewicz~disjunction)} \\
   x \land y &:= \lnot (\lnot x \lor \lnot y) &
        \text{(\L ukasiewicz~conjunction)} \\
   x \leftrightarrow_L y &:= (x \to_L y) \land (y \to_L x) &
        \text{(\L ukasiewicz~equivalence)}
\end{align*}
the former two being the algebraic realizations of the logical connectives OR and AND respectively.

From these definitions it is easy to see that the following equalities hold:
\begin{equation*}
   x \lor y = \max\{x,y\} \quad\text{and}\quad
   x \land y = \min\{x,y\}
\end{equation*}
where $\max$ and $\min$ are the lub and glb of the pair of numbers $x,y$ 
with respect to the standard total order of $L_\alpha$, which can also be expressed in the form:
\begin{equation*}
x\le y\quad\text{iff}\quad x\to_L y=1
\end{equation*}

One important feature of all many--valued connectives now presented is that
they are equal to the analogous Boolean connectives when only $0$ or $1$
are involved.

Zawirski in~\cite{zawirski} for the first time considered as primitive connective on
$L_{\alpha}$ (instead of the \L ukasiewicz implication) the binary operation of {\em
truncated sum\/} defined as follows:

\begin{align*}
   x \oplus y :&= \min\{1, x + y\} & \text{(truncated\ sum)}\\
               &=\begin{cases}
                   x + y & \text{if \ $x+y < 1$} \\
                   1         & \text{otherwise}
                 \end{cases}
\end{align*}

The two systems based on the numerical set of truth values $L_{\alpha}$, the \L
ukasiewicz one $\para{L_{\alpha},\to_L}$ and the Zawirski one $\para{L_{\alpha},\oplus}$,
are mutually equivalent owing to the ``translation'' rules:
\begin{equation}
\label{eq:mv-luk}
 x\oplus y =\lnot x\to_L y\quad\text{and}\quad x\to_L y=\neg x\oplus y
\end{equation}
Furthermore, the following binary operation can be defined in the Zawirski
$\para{L_\alpha,\oplus}$--system:
\begin{align*}
   x \odot y :&= \lnot(\lnot x\oplus\lnot y)
               =\max\{0, x + y - 1\} \\
              &=\begin{cases}
                   x + y -1& \text{if \ $1<x+y$} \\
                   0         & \text{otherwise}
                 \end{cases}
\end{align*}
In some semantical interpretations, $\oplus$ and $\odot$ are considered as algebraic realizations
of the logical connectives VEL and ET respectively, and they are also called the {\em disjunction}
and {\em conjunction} MV--connectives.

Let us stress that on the basis either of the \L ukasiewicz system or of the Zawirski one it is
always possible to derive a structure $\para{L_\alpha,\land,\lor,\neg}$ of distributive lattice
with a nonstandard negation. The lattice join and meet operations, algebraic realizations of the
logical connectives OR and AND, can be defined in the two systems respectively as follows:
\begin{subequations}
\begin{align}
   x \lor y  & = \max\{x , y\} = (x \odot \lnot y) \oplus y = \lnot (\lnot x \oplus y) \oplus y \\
   x \land y & = \min\{x , y\} = (x \oplus \lnot y) \odot y = \lnot [\lnot (x \oplus \lnot y)\oplus \lnot y]
\end{align}
\end{subequations}
Note that the {\em excluded middle law\/} holds in the case of the VEL connective
($\forall x\in L_\alpha$: $x\oplus\neg x=1$), whereas in general this law does not hold
for the OR connective ($\forall x\in L_\alpha\setminus\{0,1\}$: $x\lor\neg x\neq 1$). A
similar result is verified with respect to the {\em non--contradiction law\/} ($\forall
x\in L_\alpha$: $x\odot\neg x=0$ and $\forall x\in L_\alpha\setminus\{0,1\}$: $x\land\neg
x\neq 0$). However, the desirable law $x\lor x\to_L x=1$ holds relatively to the OR
connective, but for every $x\neq 0,1$ one has that $x\oplus x\to_L x\neq 1$.
In the Zawirski context, the standard ordering on $L_\alpha$ assumes now the form:
\begin{equation*}
x\le y\quad\text{iff}\quad \lnot x\oplus y=1
\end{equation*}

Two modal connectives, \emph{possibility} ($\pos$) and \emph{necessity} ($\nec$), can be
introduced on $L_\alpha$ according to the following definitions:
\begin{align*}
   \pos x &= \begin{cases}
                0 & \text{if \ $x = 0$} \\
                1 & \text{if \ $x \neq 0$}
             \end{cases} & \text{(possibility)} \\
   \nec x &= \begin{cases}
                0 & \text{if \ $x \neq 1$} \\
                1 & \text{if \ $x = 1$}
             \end{cases} & \text{(necessity)}
\end{align*}

Note that  these two modal connectives are mutually interdefinable, owing to the following
relationships:

\begin{equation}
\nec x = \lnot \pos \lnot x\quad \text{and}\quad \pos x = \lnot \nec \lnot x
\end{equation}
Moreover, the restriction of both connectives to the Boolean values coincides with the identity
function (these modalities are meaningless in the Boolean environment).


Besides the diametrical negation ($\lnot$) two other negation connectives can be defined as
many--valued extensions of the standard Boolean negation: the \emph{intuitionistic negation} (also
{\em impossibility\/} $\sim$) and the \emph{anti--intuitionistic negation} (also {\em
contingency\/} $\flat$) defined as:
\begin{align*}
   \sim  x &:= \neg \pos x & \text{(impossibility)} \\
           &= \begin{cases}
                 1 & \text{if \ $x = 0$} \\
                 0 & \text{if \ $x \neq 0$}
              \end{cases}   & \text{(intuitionistic~negation)}
\intertext{and}
   \flat x &:= \neg \nec x & \text{(contingency)} \\
           &= \begin{cases}
                  1 & \text{if \ $x \neq 1$} \\
                  0 & \text{if \ $x = 1$}
               \end{cases} & \text{(anti--intuitionistic~negation)}
\end{align*}
In agreement with the intuitionistic propositional logic of Brouwer and Heyting, the intuitionistic
negation ``impossibility'' fails the excluded middle law ($\forall x\in L_\alpha\setminus\{0,1\}$:
$x\lor\sim x=x\oplus \sim x=x\neq 1$), but does not fail the law of noncontradiction ($\forall x\in
L_\alpha$: $x\land\sim x=x\odot\sim x=0$). Note that the restriction of the three negations to the
two Boolean values collapses in a unique (standard) negation ($\forall x\in\{0,1\}$: $\lnot x=\sim
x=\flat x=1-x$). Trivially, also these two negation connectives are mutually interdefinable
according to:
\begin{equation}
\flat x = \lnot \sim \lnot x\quad \text{and}\quad \sim x = \lnot \flat \lnot x
\end{equation}

The intuitionistic negation is a primitive one, together with the dia\-me\-tri\-cal negation, in
\emph{BZ--lattice structures\/}, of which the system $\para{L_\alpha,\land,\lor,\neg,\sim}$ is a
standard model. Also in this case, the modal connectives can be recovered from the two involved
negations according to the following:
\begin{equation}
\nec x=\neg\sim x\quad\text{and}\quad \pos x = \sim \neg x
\end{equation}
For further information on BZ structures see \cite{cattaneo-nistico}.

In conclusion, in the algebraic approach to many--valued logics we have considered as primitive two
mutually interdefinable (according to \eqref{eq:mv-luk}) systems, the \L ukasiewicz one
$\para{L_\alpha,\to_L}$ and the Zawirski one $\para{L_\alpha,\oplus}$. An new system of
distributive lattice with diametrical negation $\para{L_\alpha,\land,\lor,\neg}$ can always be
induced. Moreover, the set of unary connectives $\{\nec,\pos,\sim,\flat\}$ (two modalities and two
negations) are mutually interdefinable making use of the diametrical negation ($\neg$) according to
the following diagram:

$$\xymatrix{
&    \pos
    \ar@{<->}[d]_{\neg(\,\cdot\,)}
    \ar@{<->}[rr]^{\neg(\,\cdot\,)\neg}
&   & \nec
    \ar@{<->}[d]^{\neg(\,\cdot\,)}
       \\
&     \sim
     \ar@{<->}[rr]_{\neg(\,\cdot\,)\neg}
&   & \flat
   }
$$
\subsubsection{The finite--valued case}
In the three--valued logic $L_3$ one has:
\begin{align*}
\pos x &= \lnot x \to_L x = x\oplus x \\
\nec x &= \lnot (x\to_L\lnot x)=\neg(\neg x\oplus \neg x)=x\odot x
\end{align*}
Let us stress that in the three valued case $L_3$ the above definition of ``it is possible that
$x$'' coincides with ``if not $x$ then $x$'' ($\pos x = \lnot x\to_L x$); in \cite{luk-30} \L
ukasiewicz mentioned that Tarski, a student of him, in 1921 proposed this as the definition of
possibility.
Therefore in this particular case we can derive the modal connectives from the system
$\para{L_3,\to_L}$ (equivalently, $\para{L_3,\oplus}$) which is thus sufficient to generate all
the connectives introduced above. In particular, the two BZ negations have the form $\sim x=\lnot(
\lnot x \to_L x )=\neg(x\oplus x)$ and $\flat x=x\to_L\lnot x=\neg x\oplus \neg x$.

In the more general finite $d$--valued case the link between possibility and VEL
connectives is extended by the following identity which is true for every $x\in L_d$:
\begin{equation*}
   \pos x := \underbrace{x \oplus x \oplus \ldots \oplus x}_{(d-1)-{\rm times}}
\end{equation*}
Thus owing to this result and the relation
\begin{equation*}
   \nec x=\neg\pos\neg x= \underbrace{x \odot x \odot \ldots \odot x}_{(d-1)-{\rm times}}
\end{equation*}
also in any finite--valued case the modal connectives of possibility and necessity can be
both derived inside the system $\para{L_d,\oplus}$.

We observe that, for infinite--valued logics, it is not possible to derive from $\to_L$ and $\lnot$
the modal operators $\nec$ and $\pos$ and the intuitionistic and anti--intuitionistic negations
$\sim$ and $\flat$ as we have just done for the finite--valued case. In fact, in \cite{mcnaughton}
it has been proved the following theorem.

\begin{theorem}
   Let $L \in \{L_{\aleph_0}, L_{\aleph_1}\}$.
   A function $f: L^m \to L$ is expressible as a formula containing only the
   operators $\to_L$ and $\lnot$ if and only if it is continuous.
\end{theorem}
\subsection{G\"odel approach}
The extension of classical connectives to many--valued logics is not uni\-que.
For example, different types of implications have been defined in literature;
one of these, which is often used, is the implication $\to_G$ defined by
G\"odel:
\begin{align*}
   x \to_G y      
             &:= \begin{cases}
                    y & \text{if \ $y < x$} \\
                    1 & \text{otherwise}
                 \end{cases} & \text{(G\"odel~implication)}
\end{align*}
Note that, if the use of the constant value 0 is allowed, we can obtain the
intuitionistic negation as $\sim x = x \to_G 0$. Moreover, in the three--valued case
G\"odel's implication differs from $\to_L$ only for the input pair $(\um, 0)$: in fact,
$\um \to_L 0 = \um$ whereas $\um \to_G 0 = 0$.
\section{Functional Completeness of Finite--valued Calculus}
%
We face now the problem whether any conceivable function $f : L_d^n \to L_d$, for $n$
ranging in $\IN$, is constructible using only the operators $\lnot$ and $\to_L$, i.e.,
the {\em functional completeness problem\/} on $L_d$ of the pair of connectives
$\{\lnot,\to_L\}$.

The following result, originally due to Jerzy S\l upecki (see, for example,
\cite{rosser-turquette}), gives a negative answer.

\begin{theorem}
   The $d$--valued (with $d \ge 3$) propositional calculus of \L ukasie\-wicz
   based on operators $\lnot$ and $\to_L$ is not functionally complete.
   That is, there exist functions $f : L_d^n  \to L_d$ which are not
   expressible as a composition of the logical functions $\lnot$ and $\to_L$ (from which we
   stress that it is possible to derive the logical functions $\lor$, $\land$, $\oplus$,
   $\odot$, $\pos$, $\nec$, $\flat$, $\sim$, $\leftrightarrow_L$).
\end{theorem}
\begin{proof}
The result follows directly from the fact that every function built up using only $\lnot$
and $\to_L$ gives a result in $\{0,1\}$ when its arguments are assigned with values in
this set. As a consequence we cannot represent, for example, the constant function which
is identically equal to $\frac{1}{d-1}$.
\end{proof}

To make the $d$--valued sentential calculus {\em functionally complete\/}
S\l upecki introduced a new unary connective, called \emph{tertium}, which is defined by the constant function ${\cal T}:L_d\to L_d$:
\begin{equation*}
   \forall \, x \in L_d \qquad {\cal T}(x) := \frac{1}{d-1}
\end{equation*}

In fact, the following theorem holds.
\begin{theorem}
   The $d$--valued (with $d \ge 2$) propositional calculus of \L ukasie\-wicz
   is functionally complete with respect to the set of
   primitive truth functions $\{\lnot,\to_L,{\cal T}\}$.
\end{theorem}

The proof of this theorem is constructive and uses the following $d$ functions:
\begin{equation*}
   j_k : L_d \to L_d, \qquad \text{for $k \in L_d$}
\end{equation*}
defined as:
\begin{equation*}
   j_k(x) = \begin{cases}
               1 & \text{if \ $x = k$} \\
               0 & \text{otherwise}
            \end{cases}
\end{equation*}

\noindent
We do not give here the proof of the theorem above mentioned due to its length. The interested
reader can find it, for example, in \cite{rosser-turquette}.
Observe that functions $j_k$ can also be defined as:
\begin{equation}
\label{eq:jk}
j_k(x) = \nec(x\leftrightarrow_L k)
\end{equation}
where the constant values $k\in L_d$ are directly involved (besides the connectives
$\nec$ and $\leftrightarrow_L$, both definable using the pair $\{\lnot,\to_L\}$).
Using these functions $\{j_k:L_d\to L_d\;|\;k\in L_d\}$ it is immediate to obtain the
tertium function ${\cal T}$ as follows:
\begin{equation}
\label{eq:ter-const}
{\cal T}(x)=\frac{1}{d-1}\sum_{k\in L_d} j_k(x)
\end{equation}

On the other hand, the $d$--valued ($d\ge 3$) propositional calculus of
\L ukasiewicz with the tertium function $\{\lnot,\to_L,{\cal T}\}$ is
functionally complete, and thus owing to~\eqref{eq:jk}
and~\eqref{eq:ter-const} also $\para{L_d,\to_L}$ is functionally complete
(recall that $\lnot x=x\to_L 0$) according to the following definition:
\begin{itemize}
\item
A collection of primitive truth functions $L_d^n\to L_d$ and a set of constants from
$L_d$ is {\em universal\/} or (according to \cite{rosser-turquette}) {\em functionally
complete\/} if and only if all possible truth functions $L_d^n\to L_d$, with $n\in\IN$,
are constructible by combining these primitive functions and assigned constants.
\end{itemize}

This means that it is functionally equivalent to assume the tertium function
or the presence of constants to the original set of primitives $\{\lnot,\to_L\}$.
\subsection{Generalization of the disjunctive normal form (GDNF)}
Let us consider the following function from $L_d^{2n}$ to $L_d$:
\begin{equation*}
   M(x_1, x_2, \ldots, x_n, c_1, c_2, \ldots, c_n) :=
      \bigwedge_{i=1}^n j_{c_i}(x_i)
\end{equation*}
It is easily verified that for every possible choice of $x_1, x_2, \ldots,
x_n, c_1, c_2, \ldots, c_n$ in $L_d$ it holds:
\begin{equation*}
   M(x_1, x_2, \ldots, x_n, c_1, c_2, \ldots, c_n) =
      \begin{cases}
         1 & \text{if $\forall \, i \in \{1, \ldots, n\}$, $x_i = c_i$} \\
         0 & \text{otherwise}
      \end{cases}
\end{equation*}

We can thus state the following theorem.
\begin{theorem}[Generalization of the disjunctive normal form]
   Let $f: L_d^n \to L_d$ be a function.
   For every choice of $(x_1, \ldots, x_n)$ in $L_d^n$ it holds:
   \begin{equation}
      f(x_1, \ldots, x_n) = \bigvee_{f(c_1, \ldots, c_n) \neq 0}
           [M(x_1, \ldots, x_n, c_1, \ldots, c_n) \wedge f(c_1, \ldots, c_n)]
        \label{eq:GDNF}
   \end{equation}
\end{theorem}

Another generalization of the disjunctive normal form which is analogous to
the one presented here can be found in \cite{urquhart}(Lemma 2.9).

Notice that if we let $E = \{0,1\}$ then expression \eqref{eq:GDNF} can be
expanded as:
\begin{align*}
   f(x_1, \ldots, x_n) = & \bigvee_{f(c_1, \ldots, c_n) \not\in E}
        [M(x_1, \ldots, x_n, c_1, \ldots, c_n) \wedge f(c_1, \ldots, c_n)]
        \lor \\
   & \bigvee_{f(c_1, \ldots, c_n) = 1}M(x_1, \ldots, x_n, c_1, \ldots, c_n)
\end{align*}
and eventually simplified by observing that $\nec (x \leftrightarrow_L 1) =
\nec x$ and $\nec (x \leftrightarrow_L 0) = \sim x$.

\subsection{Generalization of the conjunctive normal form (GCNF)}

Analogously to the disjunctive form, we introduce the function $S : L_d^{2n}
\to L_d$ defined as:
\begin{equation*}
   S(x_1, \ldots, x_n, c_1, \ldots, c_n) := \bigvee_{i=1}^n h_{c_i}(x_i)
\end{equation*}
where
\begin{equation*}
   h_k(x) := \flat (x \leftrightarrow_L k) \qquad \text{for $k \in L_d$}
\end{equation*}

The following lemma holds.
\begin{lemma}
   For every choice of $x_1, \ldots, x_n, c_1, \ldots, c_n$ in $L_d$ it holds:
   \begin{equation*}
      S(x_1, \ldots, x_n, c_1, \ldots, c_n) =
         \begin{cases}
            1 & \text{if \ $\exists \, i \in \{1, \ldots, n\}$ s.t. $x_i \neq
                      c_i$} \\
            0 & \text{if \ $\forall \, i \in \{1, \ldots, n\}$, $x_i = c_i$}
         \end{cases}
   \end{equation*}
\end{lemma}
\begin{proof}
Since $h_k(x) = \flat (x \leftrightarrow_L k) = \lnot \nec (x
\leftrightarrow_L k) = \lnot j_k(x)$, it holds:
\begin{equation*}
   h_k(x) = \begin{cases}
               1 & \text{if \ $x \neq k$} \\
               0 & \text{if \ $x = k$}
            \end{cases}
\end{equation*}
The proof of the claim follows immediately from the fact that $S$ is a
disjunction of the functions $h_k(x)$.
\end{proof}

We can thus state the following theorem.
\begin{theorem}[Generalization of the conjunctive normal form]
   Let $f: L_d^n \to L_d$ be a function.
   For every choice of $(x_1, \ldots, x_n)$ in $L_d^n$ it holds:
   \begin{equation}
      f(x_1, \ldots, x_n) = \bigwedge_{f(c_1, \ldots, c_n) \neq 1}
           [S(x_1, \ldots, x_n, c_1, \ldots, c_n) \lor f(c_1, \ldots, c_n)]
      \label{eq:GCNF}
   \end{equation}
\end{theorem}

Notice that if we let $E = \{0,1\}$ then expression \eqref{eq:GCNF} can be
expanded as:
\begin{align*}
   f(x_1, \ldots, x_n) = & \bigwedge_{f(c_1, \ldots, c_n) \not\in E}
        [S(x_1, \ldots, x_n, c_1, \ldots, c_n) \lor f(c_1, \ldots, c_n)]
        \land \\
   & \bigwedge_{f(c_1, \ldots, c_n) = 0}S(x_1, \ldots, x_n, c_1, \ldots, c_n)
\end{align*}
and eventually simplified by observing that $\flat (x \leftrightarrow_L 1) =
\flat x$ and $\flat (x \leftrightarrow_L 0) = \pos x$.

\subsection{Clay's representation}

Another way to represent the functions from $L_d^n$ to $L_d$ is given in the
following theorem, taken from \cite{clay}.
\begin{theorem}
   Let $f : L_d^n \to L_d$ be a function.
   For every choice of $(x_1, \ldots, x_n)$ in $L_d^n$ it holds:
   \begin{equation}
      f(x_1, \ldots, x_n) = \bigwedge_{f(c_1, \ldots, c_n) \neq 1}
           [M(x_1, \ldots, x_n, c_1, \ldots, c_n) \to_L f(c_1, \ldots, c_n)]
      \label{eq:Clay-representation}
   \end{equation}
\end{theorem}
\begin{proof}
It suffices to notice that, from the definition of $M$, for every choice of
$x_1, \ldots, x_n$, $c_1, \ldots, c_n$ and $y$ in $L_d$ it holds:
\begin{equation*}
   M(x_1, \ldots, x_n, c_1, \ldots, c_n) \to_L y =
      \begin{cases}
         y & \text{if \ $\forall \, i \in \{1, \ldots, n\}$, $x_i = c_i$} \\
         1 & \text{otherwise}
      \end{cases}
\end{equation*}
\end{proof}

Since $x \to_L 0 = \lnot x$, expression \eqref{eq:Clay-representation} can be
simplified as:
\begin{align*}
   f(x_1, \ldots, x_n) = & \bigwedge_{f(c_1, \ldots, c_n) \not\in E}
        [M(x_1, \ldots, x_n, c_1, \ldots, c_n) \to_L f(c_1, \ldots, c_n)]
        \land \\
   & \bigwedge_{f(c_1, \ldots, c_n) = 0} \lnot M(x_1, \ldots, x_n, c_1,
        \ldots, c_n)
\end{align*}
or, alternatively, as:
\begin{align*}
   f(x_1, \ldots, x_n) = & \bigwedge_{f(c_1, \ldots, c_n) \not\in E}
        [M(x_1, \ldots, x_n, c_1, \ldots, c_n) \to_L f(c_1, \ldots, c_n)]
        \land \\
   & \bigwedge_{f(c_1, \ldots, c_n) = 0}S(x_1, \ldots, x_n, c_1, \ldots, c_n)
\end{align*}
\section{Finite--valued Conservative Logics}
In this section we extend conservative logic to include the main features of $d$--valued
logics, with a particular attention towards three--valued logics. Since conservative
logic is based on the Fredkin gate, we will extend it in order to deal with $d$
possible truth values on its input and output lines.

First of all we restrict our attention to gates having the same number of input and output lines.
For brevity, we denote by $(n, d)$--gate an $n$--inputs/$n$--outputs gate whose input and output
lines may assume values from $L_d$. Thus, an $(n, d)$--gate computes a function $f : L_d^n \to
L_d^n$, where $L_d^n = \underbrace{L_d \times \ldots \times L_d}_{n \; {\rm times}}$. Any finite
sequence ({\em string\/}) $(x_1,x_2,\ldots,x_n)\in L^n_d$ is called a {\em configuration\/} or {\em
pattern\/} of size $n$.

{\bf Reversibility}\quad The extension of the \emph{reversibility} property is simple: an
$(n, d)$--gate is reversible if and only if the function computed by the gate is
one--to--one (or, in other words, a permutation of the set $L_d^n$). A similar argument
holds for \emph{self--reversibility}: a gate is self--reversible if and only if the
corresponding function applied twice is the identity function. As noted above, this
happens if and only if the function is a permutation which can be expressed as the
composition of disjoint cycles of length two (plus, possibly, some fixed points).

{\bf Conservativeness}\quad
More complex is the case of \emph{conservativeness}.
A gate is \emph{strictly conservative} if and only if each output configuration is a permutation of the input one.
This definition reflects perfectly the observation made by Fredkin and Toffoli
in \cite{fredkin-toffoli}, cited above on page \pageref{cit:fredkin-toffoli}.

Notice that the permutation of the input values is not fixed, but varies
depending on the pattern of values presented to the input lines; an example
can be seen in Figure \ref{fig:conditional-switch}, where two possible
permutations are chosen according to the value fed to the first input of the
Fredkin gate.

Clearly the two--valued Fredkin gate is strictly conservative, and in our first efforts
to make an extension of this gate to the finite--valued case we tried to preserve this
property. Unfortunately, if the number $n$ of input/output lines of a strictly
conservative gate for a $d$--valued logic is not greater than $d$, then it is impossible
to realize in its configurations the FAN--OUT function, as stated in the following
proposition.

\begin{proposition}
   If $n$ and $d$ are two integer numbers such that $0 < n \le d$ then there
   is no function $f: L_d^n \to L_d^n$ which corresponds to a strictly
   conservative gate realizing in its configurations the {\rm FAN--OUT} gate.
   \label{prop:noFANOUT}
\end{proposition}
\begin{proof}
If $n = 1$ then the gate has one output, and thus it cannot realize the FAN--OUT
function. So, assume that $1< n \le d$, and that there exists a strictly conservative
gate realizing FAN--OUT and corresponding to a function $f: L_d^n \to L_d^n$. In the gate
configuration realizing the FAN--OUT function, one input line is fed with a variable
value and $n-1$ input lines are fed with constant values. Since $n-1\le d-1$, there
exists at least one truth value $\ell \in L_d$ which does not appear in the fixed
constant input values. When the variable value of the input is set to $\ell$, both the
following properties should hold:
\begin{itemize}
   \item the output configuration should be a permutation of the input configuration (since the
         gate is strictly conservative), and
   \item $\ell$ should appear twice in the output values (as the gate realizes
         the FAN--OUT function),
\end{itemize}
which is clearly impossible.
\end{proof}

If the condition $n \le d$ in Proposition \ref{prop:noFANOUT} is relaxed, then it is not
difficult to see that FAN--OUT can be realized through gates which are both reversible
and strictly conservative: see, for example, the Fredkin gate, where $n = 3$ and $d = 2$.

{\bf Weak Conservativeness}\quad
An alternative approach is to weaken the conservativeness property in order to
obtain some reasonable gate that computes the FAN--OUT function.
Thus we say that a gate is \emph{weakly conservative} if and only if the sum
of output values is always equal to the sum of input values.
It is clear that if a gate is strictly conservative then it is also weakly
conservative, while the converse is not generally true.

For example, if the input of a gate is $(\lambda, 0, 1)$ and the corresponding
output is $(0, 1, \lambda)$ then the gate is both strictly conservative and
weakly conservative for this input/output pair, regardless of the numerical
value associated to $\lambda\in L_d$.
On the other hand, if the corresponding output is $(\lambda, \lambda,
\lambda)$ then the gate is weakly conservative if and only if we associate to
$\lambda$ the numerical value $\frac{1}{2}$, while it is not strictly
conservative, whatever is the numerical value associated to $\lambda$.
Indeed it is easy to see that, for a given pattern of input values, the set
of admissible patterns for output prescribed by the weak conservativeness
property varies depending upon the numerical values associated to the truth
values.

Assuming $L_d$ as the set of truth values, we propose a possible physical
interpretation of the weak conservativeness property.
To produce a given pattern of input values for a gate we need some amount of
energy.
A ``conservative'' gate has to build the output pattern in such a way that
this energy is preserved; in other words, the output produced must have the
property that, if built from scratch, it requires the same amount of energy
which was required to build the input.
The simplest way to satisfy this property is to produce a permutation of the
input values, as strictly conservative gates do.

Now, let us suppose to encode the $d$ truth values on a physical system which
has the energy levels that are equally spaced and ordered according to the
numerical value associated to the truth values.
Thus, to switch from a given truth value, say $\frac{k}{d-1}$, to the next,
that is $\frac{k+1}{d-1}$, we need to provide a fixed amount $\Delta E$ of
energy.
Analogously, when passing from a given truth value to the previous, the same
amount $\Delta E$ of energy is released.

For a gate to be conservative, it must build the output pattern without
requiring energy from an external source nor dissipating energy towards the
environment; this means that it can switch a line from a truth value
$\frac{k_1}{d-1}$ to a higher value $\frac{k_2}{d-1}$ if and only if the
energy needed (which is equal to $(k_2 - k_1) \cdot \Delta E$) becomes
available by lowering of the same amount the truth value stored in some other
line.
This is clearly equivalent to requiring that the sum of the values on the
output lines be equal to the sum of the values on the input lines.

{\bf 0 and 1--Regularity}\quad
We now define two other properties of the Fredkin gate.
They are not fundamental properties but characterize, for $d$--valued logics,
three--inputs/three--outputs gates that have a behavior which is similar to
the two--valued Fredkin gate.
We recall that the Fredkin gate exchanges the second input with the third one
when the first input is set to $0$, and it gives as outputs the inputs
unchanged when the first input is set to $1$.
According to this point of view, let $G: L_d^3 \to L_d^3$ be the function
computed by a $(3, d)$--gate; we say that the gate is \emph{$0$--regular} if
and only if $G(0, x_2, x_3) = (0, x_3, x_2)$ for every possible choice of
$x_2, x_3$ in $L_d$.
Analogously, we say that the gate is \emph{$1$--regular} if and only if
$G(1, x_2, x_3) = (1, x_2, x_3)$ for every possible choice of $x_2, x_3$ in
$L_d$.

{\bf Functional Completeness}\quad The last fundamental property satisfied by the Fredkin gate is
{\it universality} (or {\em functional completeness\/}). Indeed, according to the definition given
above, with the $d$ valued extensions of the Fredkin gate we will discuss in the next sections it
is possible to realize two universal sets for $d$--valued logics, either $ \{ \neg,\to_L, {\cal T}
\}$ or $\{ \lnot ,\oplus,{\cal T} \}$.

{\bf Conclusions}\quad
In the next sections we look for universal gates for $d$--valued logics which
preserve as many of the following properties as possible:
\begin{lquote}
\begin{itemize}
   \item[F-1~)] it is a $(3, d)$--gate, that is a three--inputs/three--outputs
                gate where each input and each output line may assume one of
                the values in $L_d = \logici$;
   \item[F-2~)] it is reversible;
   \item[F-2')] it is self--reversible;
   \item[F-3~)] it is weakly--conservative;
   \item[F-3')] it is strictly--conservative;
   \item[F-4~)] it is a universal gate, that is, from the configurations of
                the gate a universal set of connectives is obtained, included
                FAN--OUT;
   \item[F-5~)] it is $0$--regular;
   \item[F-6~)] it is $1$--regular;
   \item[F-7~)] $y_1 = x_1$, that is, the first output is always equal to the
                first input (conditional control condition);
   \item[F-8~)] when feeded with Boolean input triples it behaves as the classical Fredkin gate.
\end{itemize}
\end{lquote}
Properties F-5)--F-8) are not essential from the point of view of conservative logic, but
nonetheless are desirable, since they [characterize] the Fredkin gate.

\section{Three--valued Universal Gates}

In order to devise a universal gate for a three--valued logic, the first idea that comes
to mind is to take the equations which define the input/output behavior of the Fredkin
gate and to interpret $\lnot$, $\vee$ and $\wedge$ respectively as the \L ukasiewicz
negation, disjunction and conjunction. However this approach does not work, as the
mapping from $L_3^3$ to $L_3^3$ thus obtained is not even a bijection. As a consequence,
we have to look for gates which are universal and preserve as many properties from F-1)
-- F-8) as possible.

The next table presents all the binary three--valued connectives that we are interested to realize
with our three--valued universal gates: the \L ukasie\-wicz implication $\to_L$, the G\"odel
implication $\to_G$, the \L ukasiewicz disjunction $\lor$, the \L ukasiewicz conjunction $\land$,
the VEL--disjunction $\oplus$ and the ET--conjunction $\odot$:

\begin{center}
\begin{tabular}{|c|c|c|c|c|c|c|c|}
   \hline
   $x$ & $y$ & $\to_L$ & $\to_G$ & $\land$ & $\lor$ & $\oplus$ & $\odot$ \\
   \hline \hline
   \mbox{\rule[0cm]{0cm}{2.5ex}$0$} & $0$ & $1$ & $1$ & $0$ & $0$ & $0$ & $0$ \\
   \mbox{\rule[0cm]{0cm}{2.5ex}$0$} & $\um$ & $1$ & $1$ & $0$ & $\um$ & $\um$ & $0$ \\
   \mbox{\rule[0cm]{0cm}{2.5ex}$0$} & $1$ & $1$ & $1$ & $0$ & $1$ & $1$ & $0$ \\
   \mbox{\rule[0cm]{0cm}{2.5ex}$\um$} & $0$ & $\um$ & $0$ & $0$ & $\um$ & $\um$ & $0$ \\
   \mbox{\rule[0cm]{0cm}{2.5ex}$\um$} & $\um$ & $1$ & $1$ & $\um$ & $\um$ & $1$ & $0$ \\
   \mbox{\rule[0cm]{0cm}{2.5ex}$\um$} & $1$ & $1$ & $1$ & $\um$ & $1$ & $1$ & $\um$ \\
   \mbox{\rule[0cm]{0cm}{2.5ex}$1$} & $0$ & $0$ & $0$ & $0$ & $1$ & $1$ & $0$ \\
   \mbox{\rule[0cm]{0cm}{2.5ex}$1$} & $\um$ & $\um$ & $\um$ & $\um$ & $1$ & $1$ & $\um$ \\
   \mbox{\rule[0cm]{0cm}{2.5ex}$1$} & $1$ & $1$ & $1$ & $1$ & $1$ & $1$ & $1$ \\
   \hline
\end{tabular}
\end{center}

The unary connectives here considered are, besides the trivial identity connective Id, the negation
connectives $\lnot$, $\sim$, $\flat$ and the modal connectives $\pos$ and $\nec$ depicted in the
following table:

\begin{center}
\begin{tabular}{|c|c|c|c|c|c|}
   \hline
   $x$ & $\lnot$ & $\sim$ & $\flat$ & $\pos$ & $\nec$ \\
   \hline \hline
   \mbox{\rule[0cm]{0cm}{2.5ex}$0$} & $1$ & $1$ & $1$ & $0$ & $0$ \\
   \mbox{\rule[0cm]{0cm}{2.5ex}$\um$} & $\um$ & $0$ & $1$ & $1$ & $0$ \\
   \mbox{\rule[0cm]{0cm}{2.5ex}$1$} & $0$ & $0$ & $0$ & $1$ & $1$ \\
   \hline
\end{tabular}
\end{center}

It is important to stress that besides unary and binary connectives we must
consider the FAN--OUT gate which plays a fundamental role for reversible
computations.
Due to Proposition \ref{prop:noFANOUT}, the presence of this gate forbids the
strict conservativeness of a universal $(3,3)$--gate.

The first three--valued gate that we introduce (F$_1$) allows to obtain from
its configurations all the main connectives of the \L ukasiewicz logic
$\para{L_3,\to_L}$, as well as the G\"odel implication.
The truth table of the gate is given in Table \ref{table:GateF1}; as it can
be seen, the gate is self--reversible (and thus reversible), $0$--regular and
$1$--regular.
Moreover, it satisfies properties F-7) and F-8).

\begin{table}[ht]
\begin{center}
\begin{tabular}{|ccc|}
   \hline
   \makebox[0.4cm]{$x_1$}\makebox[0.4cm]{$x_2$}\makebox[0.4cm]{$x_3$} &
   \hspace{-0.3cm}$\mapsto$\hspace{-0.3cm} &
   \makebox[0.4cm]{$y_1$}\makebox[0.4cm]{$y_2$}\makebox[0.4cm]{$y_3$} \\
   \hline \hline
   \makebox[0.4cm]{\rule[0cm]{0cm}{2.5ex}$0$}\makebox[0.4cm]{$0$}\makebox[0.4cm]{$0$} & &
   \makebox[0.4cm]{\rule[0cm]{0cm}{2.5ex}$0$}\makebox[0.4cm]{$0$}\makebox[0.4cm]{$0$} \\
   \makebox[0.4cm]{\rule[0cm]{0cm}{2.5ex}$0$}\makebox[0.4cm]{$0$}\makebox[0.4cm]{$\um$} & &
   \makebox[0.4cm]{\rule[0cm]{0cm}{2.5ex}$0$}\makebox[0.4cm]{$\um$}\makebox[0.4cm]{$0$} \\
   \makebox[0.4cm]{\rule[0cm]{0cm}{2.5ex}$0$}\makebox[0.4cm]{$0$}\makebox[0.4cm]{$1$} & &
   \makebox[0.4cm]{\rule[0cm]{0cm}{2.5ex}$0$}\makebox[0.4cm]{$1$}\makebox[0.4cm]{$0$} \\
   \makebox[0.4cm]{\rule[0cm]{0cm}{2.5ex}$0$}\makebox[0.4cm]{$\um$}\makebox[0.4cm]{$0$} & &
   \makebox[0.4cm]{\rule[0cm]{0cm}{2.5ex}$0$}\makebox[0.4cm]{$0$}\makebox[0.4cm]{$\um$} \\
   \makebox[0.4cm]{\rule[0cm]{0cm}{2.5ex}$0$}\makebox[0.4cm]{$\um$}\makebox[0.4cm]{$\um$} & &
   \makebox[0.4cm]{\rule[0cm]{0cm}{2.5ex}$0$}\makebox[0.4cm]{$\um$}\makebox[0.4cm]{$\um$} \\
   \makebox[0.4cm]{\rule[0cm]{0cm}{2.5ex}$0$}\makebox[0.4cm]{$\um$}\makebox[0.4cm]{$1$} & &
   \makebox[0.4cm]{\rule[0cm]{0cm}{2.5ex}$0$}\makebox[0.4cm]{$1$}\makebox[0.4cm]{$\um$} \\
   \makebox[0.4cm]{\rule[0cm]{0cm}{2.5ex}$0$}\makebox[0.4cm]{$1$}\makebox[0.4cm]{$0$} & &
   \makebox[0.4cm]{\rule[0cm]{0cm}{2.5ex}$0$}\makebox[0.4cm]{$0$}\makebox[0.4cm]{$1$} \\
   \makebox[0.4cm]{\rule[0cm]{0cm}{2.5ex}$0$}\makebox[0.4cm]{$1$}\makebox[0.4cm]{$\um$} & &
   \makebox[0.4cm]{\rule[0cm]{0cm}{2.5ex}$0$}\makebox[0.4cm]{$\um$}\makebox[0.4cm]{$1$} \\
   \makebox[0.4cm]{\rule[0cm]{0cm}{2.5ex}$0$}\makebox[0.4cm]{$1$}\makebox[0.4cm]{$1$} & &
   \makebox[0.4cm]{\rule[-1.5ex]{0cm}{4ex}$0$}\makebox[0.4cm]{$1$}\makebox[0.4cm]{$1$} \\
   \hline
\end{tabular}
\begin{tabular}{|ccc|}
   \hline
   \makebox[0.4cm]{$x_1$}\makebox[0.4cm]{$x_2$}\makebox[0.4cm]{$x_3$} &
   \hspace{-0.3cm}$\mapsto$\hspace{-0.3cm} &
   \makebox[0.4cm]{$y_1$}\makebox[0.4cm]{$y_2$}\makebox[0.4cm]{$y_3$} \\
   \hline \hline
   \makebox[0.4cm]{\rule[0cm]{0cm}{2.5ex}$\um$}\makebox[0.4cm]{$0$}\makebox[0.4cm]{$0$} & &
   \makebox[0.4cm]{\rule[0cm]{0cm}{2.5ex}$\um$}\makebox[0.4cm]{$0$}\makebox[0.4cm]{$0$} \\
   \makebox[0.4cm]{\rule[0cm]{0cm}{2.5ex}$\um$}\makebox[0.4cm]{$0$}\makebox[0.4cm]{$\um$} & &
   \makebox[0.4cm]{\rule[0cm]{0cm}{2.5ex}$\um$}\makebox[0.4cm]{$0$}\makebox[0.4cm]{$\um$} \\
   \makebox[0.4cm]{\rule[0cm]{0cm}{2.5ex}$\um$}\makebox[0.4cm]{$0$}\makebox[0.4cm]{$1$} & &
   \makebox[0.4cm]{\rule[0cm]{0cm}{2.5ex}$\um$}\makebox[0.4cm]{$0$}\makebox[0.4cm]{$1$} \\
   \makebox[0.4cm]{\rule[0cm]{0cm}{2.5ex}$\um$}\makebox[0.4cm]{$\um$}\makebox[0.4cm]{$0$} & &
   \makebox[0.4cm]{\rule[0cm]{0cm}{2.5ex}$\um$}\makebox[0.4cm]{$\um$}\makebox[0.4cm]{$0$} \\
   \makebox[0.4cm]{\rule[0cm]{0cm}{2.5ex}$\um$}\makebox[0.4cm]{$\um$}\makebox[0.4cm]{$\um$} & ** &
   \makebox[0.4cm]{\rule[0cm]{0cm}{2.5ex}$\um$}\makebox[0.4cm]{$1$}\makebox[0.4cm]{$0$} \\
   \makebox[0.4cm]{\rule[0cm]{0cm}{2.5ex}$\um$}\makebox[0.4cm]{$\um$}\makebox[0.4cm]{$1$} & &
   \makebox[0.4cm]{\rule[0cm]{0cm}{2.5ex}$\um$}\makebox[0.4cm]{$1$}\makebox[0.4cm]{$\um$} \\
   \makebox[0.4cm]{\rule[0cm]{0cm}{2.5ex}$\um$}\makebox[0.4cm]{$1$}\makebox[0.4cm]{$0$} & ** &
   \makebox[0.4cm]{\rule[0cm]{0cm}{2.5ex}$\um$}\makebox[0.4cm]{$\um$}\makebox[0.4cm]{$\um$} \\
   \makebox[0.4cm]{\rule[0cm]{0cm}{2.5ex}$\um$}\makebox[0.4cm]{$1$}\makebox[0.4cm]{$\um$} & &
   \makebox[0.4cm]{\rule[0cm]{0cm}{2.5ex}$\um$}\makebox[0.4cm]{$\um$}\makebox[0.4cm]{$1$} \\
   \makebox[0.4cm]{\rule[0cm]{0cm}{2.5ex}$\um$}\makebox[0.4cm]{$1$}\makebox[0.4cm]{$1$} & &
   \makebox[0.4cm]{\rule[-1.5ex]{0cm}{4ex}$\um$}\makebox[0.4cm]{$1$}\makebox[0.4cm]{$1$} \\
   \hline
\end{tabular}
\begin{tabular}{|ccc|}
   \hline
   \makebox[0.4cm]{$x_1$}\makebox[0.4cm]{$x_2$}\makebox[0.4cm]{$x_3$} &
   \hspace{-0.3cm}$\mapsto$\hspace{-0.3cm} &
   \makebox[0.4cm]{$y_1$}\makebox[0.4cm]{$y_2$}\makebox[0.4cm]{$y_3$} \\
   \hline \hline
   \makebox[0.4cm]{\rule[0cm]{0cm}{2.5ex}$1$}\makebox[0.4cm]{$0$}\makebox[0.4cm]{$0$} & &
   \makebox[0.4cm]{\rule[0cm]{0cm}{2.5ex}$1$}\makebox[0.4cm]{$0$}\makebox[0.4cm]{$0$} \\
   \makebox[0.4cm]{\rule[0cm]{0cm}{2.5ex}$1$}\makebox[0.4cm]{$0$}\makebox[0.4cm]{$\um$} & &
   \makebox[0.4cm]{\rule[0cm]{0cm}{2.5ex}$1$}\makebox[0.4cm]{$0$}\makebox[0.4cm]{$\um$} \\
   \makebox[0.4cm]{\rule[0cm]{0cm}{2.5ex}$1$}\makebox[0.4cm]{$0$}\makebox[0.4cm]{$1$} & &
   \makebox[0.4cm]{\rule[0cm]{0cm}{2.5ex}$1$}\makebox[0.4cm]{$0$}\makebox[0.4cm]{$1$} \\
   \makebox[0.4cm]{\rule[0cm]{0cm}{2.5ex}$1$}\makebox[0.4cm]{$\um$}\makebox[0.4cm]{$0$} & &
   \makebox[0.4cm]{\rule[0cm]{0cm}{2.5ex}$1$}\makebox[0.4cm]{$\um$}\makebox[0.4cm]{$0$} \\
   \makebox[0.4cm]{\rule[0cm]{0cm}{2.5ex}$1$}\makebox[0.4cm]{$\um$}\makebox[0.4cm]{$\um$} & &
   \makebox[0.4cm]{\rule[0cm]{0cm}{2.5ex}$1$}\makebox[0.4cm]{$\um$}\makebox[0.4cm]{$\um$} \\
   \makebox[0.4cm]{\rule[0cm]{0cm}{2.5ex}$1$}\makebox[0.4cm]{$\um$}\makebox[0.4cm]{$1$} & &
   \makebox[0.4cm]{\rule[0cm]{0cm}{2.5ex}$1$}\makebox[0.4cm]{$\um$}\makebox[0.4cm]{$1$} \\
   \makebox[0.4cm]{\rule[0cm]{0cm}{2.5ex}$1$}\makebox[0.4cm]{$1$}\makebox[0.4cm]{$0$} & &
   \makebox[0.4cm]{\rule[0cm]{0cm}{2.5ex}$1$}\makebox[0.4cm]{$1$}\makebox[0.4cm]{$0$} \\
   \makebox[0.4cm]{\rule[0cm]{0cm}{2.5ex}$1$}\makebox[0.4cm]{$1$}\makebox[0.4cm]{$\um$} & &
   \makebox[0.4cm]{\rule[0cm]{0cm}{2.5ex}$1$}\makebox[0.4cm]{$1$}\makebox[0.4cm]{$\um$} \\
   \makebox[0.4cm]{\rule[0cm]{0cm}{2.5ex}$1$}\makebox[0.4cm]{$1$}\makebox[0.4cm]{$1$} & &
   \makebox[0.4cm]{\rule[-1.5ex]{0cm}{4ex}$1$}\makebox[0.4cm]{$1$}\makebox[0.4cm]{$1$} \\
   \hline
\end{tabular}
\caption{Truth table of gate F$_1$}
\label{table:GateF1}
\end{center}
\end{table}

Table \ref{table:F1-Operators} shows all the relevant connectives which can be obtained from the
gate by fixing one or two of its input lines with constant values from $L_3$; Pr$_1$ and Pr$_2$ are
the projectors connectives defined as ${\rm Pr}_1(x_1, x_2) = x_1$ and ${\rm Pr}_2(x_1, x_2) = x_2$
respectively. We can observe that this gate realizes two negations (the diametrical and the
intuitionistic one) and both the \L ukasiewicz and G\"odel implications introduced in Section
\ref{section:MVLogic}; as a consequence, the universality property F-4) is satisfied for both kinds
of three--valued logic. On the other hand, the necessity modal connective, the anti--intuitionistic
negation and both the binary MV--connectives are not realized.

\begin{table}[ht]
\begin{center}
\begin{tabular}{|c|c|c|c|c|}
   \hline
   Connective & Inputs & Constants & Outputs & Garbage \\
   \hline \hline
   \mbox{\rule[0cm]{0cm}{2.5ex}FAN--OUT} & $x_1$ & $x_2 = 1, x_3 = 0$ &
        $y_1, y_2$ & $y_3$ \\
   Pr$_1$  & $x_2, x_3$ & $x_1 = 0$ & $y_3$ & $y_1, y_2$ \\
   Pr$_2$  & $x_2, x_3$ & $x_1 = 0$ & $y_2$ & $y_1, y_3$ \\
   $\to_L$ & $x_1, x_3$ & $x_2 = 1$ & $y_3$ & $y_1, y_2$ \\
   $\to_G$ & $x_1, x_2$ & $x_3 = 1$ & $y_2$ & $y_1, y_3$ \\
   $\lor$  & $x_1, x_3$ & $x_2 = 1$ & $y_2$ & $y_1, y_3$ \\
   $\land$ & $x_1, x_2$ & $x_3 = 0$ & $y_2$ & $y_1, y_3$ \\
   Id      & $x_1$ & $x_2 = 0, x_3 = 0$ & $y_1$ & $y_2, y_3$ \\
   $\lnot$ & $x_1$ & $x_2 = 1, x_3 = 0$ & $y_3$ & $y_1, y_2$ \\
   $\sim$  & $x_1$ & $x_2 = 0, x_3 = 1$ & $y_2$ & $y_1, y_3$ \\
   \mbox{\rule[-1.5ex]{0cm}{1.5ex}$\pos$} & $x_1$ & $x_2 = 0, x_3 = 1$ &
         $y_3$ & $y_1, y_2$ \\
   \hline
\end{tabular}
\end{center}
\caption{The operators obtained through gate F$_1$}
\label{table:F1-Operators}
\end{table}

Due to Proposition \ref{prop:noFANOUT}, gate F$_1$ cannot be strictly conservative, as it realizes
the FAN--OUT function. More precisely, strict conservativeness is lost in the two table rows marked
with $(**)$. However, for these rows the gate is weakly conservative, and therefore the entire gate
is weakly conservative.

The next two gates that we introduce are part of the results of an exhaustive
search --- performed with a program written on purpose --- over all
three--valued gates having the following properties:
\begin{lquote}
\begin{itemize}
   \item[F-1~)] it is a $(3, 3)$--gate;
   \item[F-2')] it is self--reversible;
   \item[F-3~)] it is weakly conservative;
   \item[F-8~)] when feeded with Boolean input triples it behaves as the
                Fredkin gate.
\end{itemize}
\end{lquote}

The first of the two obtained gates (F$_2$) is substantially equivalent to F$_1$; its truth table
is given in Table \ref{table:GateF2}. As we can see, this gate differs from F$_1$ only for the
input triples $0 \, \um \, \um$ and $\um \, 0 \, \um$. It is only $1$--regular and it has not the
property F-7).

\begin{table}[ht]
\begin{center}
\begin{tabular}{|ccc|}
   \hline
   \makebox[0.4cm]{$x_1$}\makebox[0.4cm]{$x_2$}\makebox[0.4cm]{$x_3$} &
   \hspace{-0.3cm}$\mapsto$\hspace{-0.3cm} &
   \makebox[0.4cm]{$y_1$}\makebox[0.4cm]{$y_2$}\makebox[0.4cm]{$y_3$} \\
   \hline \hline
   \makebox[0.4cm]{\rule[0cm]{0cm}{2.5ex}$0$}\makebox[0.4cm]{$0$}\makebox[0.4cm]{$0$} & &
   \makebox[0.4cm]{\rule[0cm]{0cm}{2.5ex}$0$}\makebox[0.4cm]{$0$}\makebox[0.4cm]{$0$} \\
   \makebox[0.4cm]{\rule[0cm]{0cm}{2.5ex}$0$}\makebox[0.4cm]{$0$}\makebox[0.4cm]{$\um$} & &
   \makebox[0.4cm]{\rule[0cm]{0cm}{2.5ex}$0$}\makebox[0.4cm]{$\um$}\makebox[0.4cm]{$0$} \\
   \makebox[0.4cm]{\rule[0cm]{0cm}{2.5ex}$0$}\makebox[0.4cm]{$0$}\makebox[0.4cm]{$1$} & &
   \makebox[0.4cm]{\rule[0cm]{0cm}{2.5ex}$0$}\makebox[0.4cm]{$1$}\makebox[0.4cm]{$0$} \\
   \makebox[0.4cm]{\rule[0cm]{0cm}{2.5ex}$0$}\makebox[0.4cm]{$\um$}\makebox[0.4cm]{$0$} & &
   \makebox[0.4cm]{\rule[0cm]{0cm}{2.5ex}$0$}\makebox[0.4cm]{$0$}\makebox[0.4cm]{$\um$} \\
   \makebox[0.4cm]{\rule[0cm]{0cm}{2.5ex}$0$}\makebox[0.4cm]{$\um$}\makebox[0.4cm]{$\um$} & &
   \makebox[0.4cm]{\rule[0cm]{0cm}{2.5ex}$\um$}\makebox[0.4cm]{$0$}\makebox[0.4cm]{$\um$} \\
   \makebox[0.4cm]{\rule[0cm]{0cm}{2.5ex}$0$}\makebox[0.4cm]{$\um$}\makebox[0.4cm]{$1$} & &
   \makebox[0.4cm]{\rule[0cm]{0cm}{2.5ex}$0$}\makebox[0.4cm]{$1$}\makebox[0.4cm]{$\um$} \\
   \makebox[0.4cm]{\rule[0cm]{0cm}{2.5ex}$0$}\makebox[0.4cm]{$1$}\makebox[0.4cm]{$0$} & &
   \makebox[0.4cm]{\rule[0cm]{0cm}{2.5ex}$0$}\makebox[0.4cm]{$0$}\makebox[0.4cm]{$1$} \\
   \makebox[0.4cm]{\rule[0cm]{0cm}{2.5ex}$0$}\makebox[0.4cm]{$1$}\makebox[0.4cm]{$\um$} & &
   \makebox[0.4cm]{\rule[0cm]{0cm}{2.5ex}$0$}\makebox[0.4cm]{$\um$}\makebox[0.4cm]{$1$} \\
   \makebox[0.4cm]{\rule[0cm]{0cm}{2.5ex}$0$}\makebox[0.4cm]{$1$}\makebox[0.4cm]{$1$} & &
   \makebox[0.4cm]{\rule[-1.5ex]{0cm}{4ex}$0$}\makebox[0.4cm]{$1$}\makebox[0.4cm]{$1$} \\
   \hline
\end{tabular}
\begin{tabular}{|ccc|}
   \hline
   \makebox[0.4cm]{$x_1$}\makebox[0.4cm]{$x_2$}\makebox[0.4cm]{$x_3$} &
   \hspace{-0.3cm}$\mapsto$\hspace{-0.3cm} &
   \makebox[0.4cm]{$y_1$}\makebox[0.4cm]{$y_2$}\makebox[0.4cm]{$y_3$} \\
   \hline \hline
   \makebox[0.4cm]{\rule[0cm]{0cm}{2.5ex}$\um$}\makebox[0.4cm]{$0$}\makebox[0.4cm]{$0$} & &
   \makebox[0.4cm]{\rule[0cm]{0cm}{2.5ex}$\um$}\makebox[0.4cm]{$0$}\makebox[0.4cm]{$0$} \\
   \makebox[0.4cm]{\rule[0cm]{0cm}{2.5ex}$\um$}\makebox[0.4cm]{$0$}\makebox[0.4cm]{$\um$} & &
   \makebox[0.4cm]{\rule[0cm]{0cm}{2.5ex}$0$}\makebox[0.4cm]{$\um$}\makebox[0.4cm]{$\um$} \\
   \makebox[0.4cm]{\rule[0cm]{0cm}{2.5ex}$\um$}\makebox[0.4cm]{$0$}\makebox[0.4cm]{$1$} & &
   \makebox[0.4cm]{\rule[0cm]{0cm}{2.5ex}$\um$}\makebox[0.4cm]{$0$}\makebox[0.4cm]{$1$} \\
   \makebox[0.4cm]{\rule[0cm]{0cm}{2.5ex}$\um$}\makebox[0.4cm]{$\um$}\makebox[0.4cm]{$0$} & &
   \makebox[0.4cm]{\rule[0cm]{0cm}{2.5ex}$\um$}\makebox[0.4cm]{$\um$}\makebox[0.4cm]{$0$} \\
   \makebox[0.4cm]{\rule[0cm]{0cm}{2.5ex}$\um$}\makebox[0.4cm]{$\um$}\makebox[0.4cm]{$\um$} & &
   \makebox[0.4cm]{\rule[0cm]{0cm}{2.5ex}$\um$}\makebox[0.4cm]{$1$}\makebox[0.4cm]{$0$} \\
   \makebox[0.4cm]{\rule[0cm]{0cm}{2.5ex}$\um$}\makebox[0.4cm]{$\um$}\makebox[0.4cm]{$1$} & &
   \makebox[0.4cm]{\rule[0cm]{0cm}{2.5ex}$\um$}\makebox[0.4cm]{$1$}\makebox[0.4cm]{$\um$} \\
   \makebox[0.4cm]{\rule[0cm]{0cm}{2.5ex}$\um$}\makebox[0.4cm]{$1$}\makebox[0.4cm]{$0$} & &
   \makebox[0.4cm]{\rule[0cm]{0cm}{2.5ex}$\um$}\makebox[0.4cm]{$\um$}\makebox[0.4cm]{$\um$} \\
   \makebox[0.4cm]{\rule[0cm]{0cm}{2.5ex}$\um$}\makebox[0.4cm]{$1$}\makebox[0.4cm]{$\um$} & &
   \makebox[0.4cm]{\rule[0cm]{0cm}{2.5ex}$\um$}\makebox[0.4cm]{$\um$}\makebox[0.4cm]{$1$} \\
   \makebox[0.4cm]{\rule[0cm]{0cm}{2.5ex}$\um$}\makebox[0.4cm]{$1$}\makebox[0.4cm]{$1$} & &
   \makebox[0.4cm]{\rule[-1.5ex]{0cm}{4ex}$\um$}\makebox[0.4cm]{$1$}\makebox[0.4cm]{$1$} \\
   \hline
\end{tabular}
\begin{tabular}{|ccc|}
   \hline
   \makebox[0.4cm]{$x_1$}\makebox[0.4cm]{$x_2$}\makebox[0.4cm]{$x_3$} &
   \hspace{-0.3cm}$\mapsto$\hspace{-0.3cm} &
   \makebox[0.4cm]{$y_1$}\makebox[0.4cm]{$y_2$}\makebox[0.4cm]{$y_3$} \\
   \hline \hline
   \makebox[0.4cm]{\rule[0cm]{0cm}{2.5ex}$1$}\makebox[0.4cm]{$0$}\makebox[0.4cm]{$0$} & &
   \makebox[0.4cm]{\rule[0cm]{0cm}{2.5ex}$1$}\makebox[0.4cm]{$0$}\makebox[0.4cm]{$0$} \\
   \makebox[0.4cm]{\rule[0cm]{0cm}{2.5ex}$1$}\makebox[0.4cm]{$0$}\makebox[0.4cm]{$\um$} & &
   \makebox[0.4cm]{\rule[0cm]{0cm}{2.5ex}$1$}\makebox[0.4cm]{$0$}\makebox[0.4cm]{$\um$} \\
   \makebox[0.4cm]{\rule[0cm]{0cm}{2.5ex}$1$}\makebox[0.4cm]{$0$}\makebox[0.4cm]{$1$} & &
   \makebox[0.4cm]{\rule[0cm]{0cm}{2.5ex}$1$}\makebox[0.4cm]{$0$}\makebox[0.4cm]{$1$} \\
   \makebox[0.4cm]{\rule[0cm]{0cm}{2.5ex}$1$}\makebox[0.4cm]{$\um$}\makebox[0.4cm]{$0$} & &
   \makebox[0.4cm]{\rule[0cm]{0cm}{2.5ex}$1$}\makebox[0.4cm]{$\um$}\makebox[0.4cm]{$0$} \\
   \makebox[0.4cm]{\rule[0cm]{0cm}{2.5ex}$1$}\makebox[0.4cm]{$\um$}\makebox[0.4cm]{$\um$} & &
   \makebox[0.4cm]{\rule[0cm]{0cm}{2.5ex}$1$}\makebox[0.4cm]{$\um$}\makebox[0.4cm]{$\um$} \\
   \makebox[0.4cm]{\rule[0cm]{0cm}{2.5ex}$1$}\makebox[0.4cm]{$\um$}\makebox[0.4cm]{$1$} & &
   \makebox[0.4cm]{\rule[0cm]{0cm}{2.5ex}$1$}\makebox[0.4cm]{$\um$}\makebox[0.4cm]{$1$} \\
   \makebox[0.4cm]{\rule[0cm]{0cm}{2.5ex}$1$}\makebox[0.4cm]{$1$}\makebox[0.4cm]{$0$} & &
   \makebox[0.4cm]{\rule[0cm]{0cm}{2.5ex}$1$}\makebox[0.4cm]{$1$}\makebox[0.4cm]{$0$} \\
   \makebox[0.4cm]{\rule[0cm]{0cm}{2.5ex}$1$}\makebox[0.4cm]{$1$}\makebox[0.4cm]{$\um$} & &
   \makebox[0.4cm]{\rule[0cm]{0cm}{2.5ex}$1$}\makebox[0.4cm]{$1$}\makebox[0.4cm]{$\um$} \\
   \makebox[0.4cm]{\rule[0cm]{0cm}{2.5ex}$1$}\makebox[0.4cm]{$1$}\makebox[0.4cm]{$1$} & &
   \makebox[0.4cm]{\rule[-1.5ex]{0cm}{4ex}$1$}\makebox[0.4cm]{$1$}\makebox[0.4cm]{$1$} \\
   \hline
\end{tabular}
\caption{Truth table of gate F$_2$}
\label{table:GateF2}
\end{center}
\end{table}

Table \ref{table:F2-Operators} shows all the relevant connectives which can be
obtained from the gate by fixing one or two of its input lines with constant
values from $L_3$. We can observe that the set of connectives is the same as
F$_1$'s with the exception of the modal connective $\nec$, which is not
present in the first gate. Thus, the deficiencies with respect to gate F$_1$
concerning the properties enjoyed by the gate are balanced with a richer set
of realized connectives. As it does happen with gate F$_1$, the set of
connectives realized by the gate F$_2$ satisfies condition F-4) of
universality.

\begin{table}[ht]
\begin{center}
\begin{tabular}{|c|c|c|c|c|}
   \hline
   Connective & Inputs & Constants & Outputs & Garbage \\
   \hline \hline
   \mbox{\rule[0cm]{0cm}{2.5ex}FAN--OUT} & $x_1$ & $x_2 = 1, x_3 = 0$ &
        $y_1, y_2$ & $y_3$ \\
   Pr$_1$  & $x_2, x_3$ & $x_1 = 1$ & $y_2$ & $y_1, y_3$ \\
   Pr$_2$  & $x_2, x_3$ & $x_1 = 1$ & $y_3$ & $y_1, y_2$ \\
   $\to_L$ & $x_1, x_3$ & $x_2 = 1$ & $y_3$ & $y_1, y_2$ \\
   $\to_G$ & $x_1, x_2$ & $x_3 = 1$ & $y_2$ & $y_1, y_3$ \\
   $\lor$  & $x_1, x_3$ & $x_2 = 1$ & $y_2$ & $y_1, y_3$ \\
   $\land$ & $x_1, x_2$ & $x_3 = 0$ & $y_2$ & $y_1, y_3$ \\
   Id      & $x_1$ & $x_2 = 0, x_3 = 0$ & $y_1$ & $y_2, y_3$ \\
   $\lnot$ & $x_1$ & $x_2 = 1, x_3 = 0$ & $y_3$ & $y_1, y_2$ \\
   $\sim$  & $x_1$ & $x_2 = 0, x_3 = 1$ & $y_2$ & $y_1, y_3$ \\
   $\pos$  & $x_1$ & $x_2 = 0, x_3 = 1$ & $y_3$ & $y_1, y_2$ \\
   \mbox{\rule[-1.5ex]{0cm}{1.5ex}$\nec$} & $x_1$ & $x_2 = 0, x_3 = \um$ &
        $y_1$ & $y_2, y_3$ \\
   \hline
\end{tabular}
\end{center}
\caption{The operators obtained through gate F$_2$}
\label{table:F2-Operators}
\end{table}

The last gate (F$_3$) we introduce allows one to realize the MV--connectives of the $3$--valued
case; its truth table is given in Table \ref{table:GateF3}. Besides properties F-1), F-2'), F-3)
and F-8), used by our program as the criteria for the exhaustive search, this gate satisfies
property F-7) of conditional control; moreover, it is $0$--regular and $1$--regular.

\begin{table}[ht]
\begin{center}
\begin{tabular}{|ccc|}
   \hline
   \makebox[0.4cm]{$x_1$}\makebox[0.4cm]{$x_2$}\makebox[0.4cm]{$x_3$} &
   \hspace{-0.3cm}$\mapsto$\hspace{-0.3cm} &
   \makebox[0.4cm]{$y_1$}\makebox[0.4cm]{$y_2$}\makebox[0.4cm]{$y_3$} \\
   \hline \hline
   \makebox[0.4cm]{\rule[0cm]{0cm}{2.5ex}$0$}\makebox[0.4cm]{$0$}\makebox[0.4cm]{$0$} & &
   \makebox[0.4cm]{\rule[0cm]{0cm}{2.5ex}$0$}\makebox[0.4cm]{$0$}\makebox[0.4cm]{$0$} \\
   \makebox[0.4cm]{\rule[0cm]{0cm}{2.5ex}$0$}\makebox[0.4cm]{$0$}\makebox[0.4cm]{$\um$} & &
   \makebox[0.4cm]{\rule[0cm]{0cm}{2.5ex}$0$}\makebox[0.4cm]{$\um$}\makebox[0.4cm]{$0$} \\
   \makebox[0.4cm]{\rule[0cm]{0cm}{2.5ex}$0$}\makebox[0.4cm]{$0$}\makebox[0.4cm]{$1$} & &
   \makebox[0.4cm]{\rule[0cm]{0cm}{2.5ex}$0$}\makebox[0.4cm]{$1$}\makebox[0.4cm]{$0$} \\
   \makebox[0.4cm]{\rule[0cm]{0cm}{2.5ex}$0$}\makebox[0.4cm]{$\um$}\makebox[0.4cm]{$0$} & &
   \makebox[0.4cm]{\rule[0cm]{0cm}{2.5ex}$0$}\makebox[0.4cm]{$0$}\makebox[0.4cm]{$\um$} \\
   \makebox[0.4cm]{\rule[0cm]{0cm}{2.5ex}$0$}\makebox[0.4cm]{$\um$}\makebox[0.4cm]{$\um$} & &
   \makebox[0.4cm]{\rule[0cm]{0cm}{2.5ex}$0$}\makebox[0.4cm]{$\um$}\makebox[0.4cm]{$\um$} \\
   \makebox[0.4cm]{\rule[0cm]{0cm}{2.5ex}$0$}\makebox[0.4cm]{$\um$}\makebox[0.4cm]{$1$} & &
   \makebox[0.4cm]{\rule[0cm]{0cm}{2.5ex}$0$}\makebox[0.4cm]{$1$}\makebox[0.4cm]{$\um$} \\
   \makebox[0.4cm]{\rule[0cm]{0cm}{2.5ex}$0$}\makebox[0.4cm]{$1$}\makebox[0.4cm]{$0$} & &
   \makebox[0.4cm]{\rule[0cm]{0cm}{2.5ex}$0$}\makebox[0.4cm]{$0$}\makebox[0.4cm]{$1$} \\
   \makebox[0.4cm]{\rule[0cm]{0cm}{2.5ex}$0$}\makebox[0.4cm]{$1$}\makebox[0.4cm]{$\um$} & &
   \makebox[0.4cm]{\rule[0cm]{0cm}{2.5ex}$0$}\makebox[0.4cm]{$\um$}\makebox[0.4cm]{$1$} \\
   \makebox[0.4cm]{\rule[0cm]{0cm}{2.5ex}$0$}\makebox[0.4cm]{$1$}\makebox[0.4cm]{$1$} & &
   \makebox[0.4cm]{\rule[-1.5ex]{0cm}{4ex}$0$}\makebox[0.4cm]{$1$}\makebox[0.4cm]{$1$} \\
   \hline
\end{tabular}
\begin{tabular}{|ccc|}
   \hline
   \makebox[0.4cm]{$x_1$}\makebox[0.4cm]{$x_2$}\makebox[0.4cm]{$x_3$} &
   \hspace{-0.3cm}$\mapsto$\hspace{-0.3cm} &
   \makebox[0.4cm]{$y_1$}\makebox[0.4cm]{$y_2$}\makebox[0.4cm]{$y_3$} \\
   \hline \hline
   \makebox[0.4cm]{\rule[0cm]{0cm}{2.5ex}$\um$}\makebox[0.4cm]{$0$}\makebox[0.4cm]{$0$} & &
   \makebox[0.4cm]{\rule[0cm]{0cm}{2.5ex}$\um$}\makebox[0.4cm]{$0$}\makebox[0.4cm]{$0$} \\
   \makebox[0.4cm]{\rule[0cm]{0cm}{2.5ex}$\um$}\makebox[0.4cm]{$0$}\makebox[0.4cm]{$\um$} & &
   \makebox[0.4cm]{\rule[0cm]{0cm}{2.5ex}$\um$}\makebox[0.4cm]{$\um$}\makebox[0.4cm]{$0$} \\
   \makebox[0.4cm]{\rule[0cm]{0cm}{2.5ex}$\um$}\makebox[0.4cm]{$0$}\makebox[0.4cm]{$1$} & &
   \makebox[0.4cm]{\rule[0cm]{0cm}{2.5ex}$\um$}\makebox[0.4cm]{$0$}\makebox[0.4cm]{$1$} \\
   \makebox[0.4cm]{\rule[0cm]{0cm}{2.5ex}$\um$}\makebox[0.4cm]{$\um$}\makebox[0.4cm]{$0$} & &
   \makebox[0.4cm]{\rule[0cm]{0cm}{2.5ex}$\um$}\makebox[0.4cm]{$0$}\makebox[0.4cm]{$\um$} \\
   \makebox[0.4cm]{\rule[0cm]{0cm}{2.5ex}$\um$}\makebox[0.4cm]{$\um$}\makebox[0.4cm]{$\um$} & &
   \makebox[0.4cm]{\rule[0cm]{0cm}{2.5ex}$\um$}\makebox[0.4cm]{$1$}\makebox[0.4cm]{$0$} \\
   \makebox[0.4cm]{\rule[0cm]{0cm}{2.5ex}$\um$}\makebox[0.4cm]{$\um$}\makebox[0.4cm]{$1$} & &
   \makebox[0.4cm]{\rule[0cm]{0cm}{2.5ex}$\um$}\makebox[0.4cm]{$\um$}\makebox[0.4cm]{$1$} \\
   \makebox[0.4cm]{\rule[0cm]{0cm}{2.5ex}$\um$}\makebox[0.4cm]{$1$}\makebox[0.4cm]{$0$} & &
   \makebox[0.4cm]{\rule[0cm]{0cm}{2.5ex}$\um$}\makebox[0.4cm]{$\um$}\makebox[0.4cm]{$\um$} \\
   \makebox[0.4cm]{\rule[0cm]{0cm}{2.5ex}$\um$}\makebox[0.4cm]{$1$}\makebox[0.4cm]{$\um$} & &
   \makebox[0.4cm]{\rule[0cm]{0cm}{2.5ex}$\um$}\makebox[0.4cm]{$1$}\makebox[0.4cm]{$\um$} \\
   \makebox[0.4cm]{\rule[0cm]{0cm}{2.5ex}$\um$}\makebox[0.4cm]{$1$}\makebox[0.4cm]{$1$} & &
   \makebox[0.4cm]{\rule[-1.5ex]{0cm}{4ex}$\um$}\makebox[0.4cm]{$1$}\makebox[0.4cm]{$1$} \\
   \hline
\end{tabular}
\begin{tabular}{|ccc|}
   \hline
   \makebox[0.4cm]{$x_1$}\makebox[0.4cm]{$x_2$}\makebox[0.4cm]{$x_3$} &
   \hspace{-0.3cm}$\mapsto$\hspace{-0.3cm} &
   \makebox[0.4cm]{$y_1$}\makebox[0.4cm]{$y_2$}\makebox[0.4cm]{$y_3$} \\
   \hline \hline
   \makebox[0.4cm]{\rule[0cm]{0cm}{2.5ex}$1$}\makebox[0.4cm]{$0$}\makebox[0.4cm]{$0$} & &
   \makebox[0.4cm]{\rule[0cm]{0cm}{2.5ex}$1$}\makebox[0.4cm]{$0$}\makebox[0.4cm]{$0$} \\
   \makebox[0.4cm]{\rule[0cm]{0cm}{2.5ex}$1$}\makebox[0.4cm]{$0$}\makebox[0.4cm]{$\um$} & &
   \makebox[0.4cm]{\rule[0cm]{0cm}{2.5ex}$1$}\makebox[0.4cm]{$0$}\makebox[0.4cm]{$\um$} \\
   \makebox[0.4cm]{\rule[0cm]{0cm}{2.5ex}$1$}\makebox[0.4cm]{$0$}\makebox[0.4cm]{$1$} & &
   \makebox[0.4cm]{\rule[0cm]{0cm}{2.5ex}$1$}\makebox[0.4cm]{$0$}\makebox[0.4cm]{$1$} \\
   \makebox[0.4cm]{\rule[0cm]{0cm}{2.5ex}$1$}\makebox[0.4cm]{$\um$}\makebox[0.4cm]{$0$} & &
   \makebox[0.4cm]{\rule[0cm]{0cm}{2.5ex}$1$}\makebox[0.4cm]{$\um$}\makebox[0.4cm]{$0$} \\
   \makebox[0.4cm]{\rule[0cm]{0cm}{2.5ex}$1$}\makebox[0.4cm]{$\um$}\makebox[0.4cm]{$\um$} & &
   \makebox[0.4cm]{\rule[0cm]{0cm}{2.5ex}$1$}\makebox[0.4cm]{$\um$}\makebox[0.4cm]{$\um$} \\
   \makebox[0.4cm]{\rule[0cm]{0cm}{2.5ex}$1$}\makebox[0.4cm]{$\um$}\makebox[0.4cm]{$1$} & &
   \makebox[0.4cm]{\rule[0cm]{0cm}{2.5ex}$1$}\makebox[0.4cm]{$\um$}\makebox[0.4cm]{$1$} \\
   \makebox[0.4cm]{\rule[0cm]{0cm}{2.5ex}$1$}\makebox[0.4cm]{$1$}\makebox[0.4cm]{$0$} & &
   \makebox[0.4cm]{\rule[0cm]{0cm}{2.5ex}$1$}\makebox[0.4cm]{$1$}\makebox[0.4cm]{$0$} \\
   \makebox[0.4cm]{\rule[0cm]{0cm}{2.5ex}$1$}\makebox[0.4cm]{$1$}\makebox[0.4cm]{$\um$} & &
   \makebox[0.4cm]{\rule[0cm]{0cm}{2.5ex}$1$}\makebox[0.4cm]{$1$}\makebox[0.4cm]{$\um$} \\
   \makebox[0.4cm]{\rule[0cm]{0cm}{2.5ex}$1$}\makebox[0.4cm]{$1$}\makebox[0.4cm]{$1$} & &
   \makebox[0.4cm]{\rule[-1.5ex]{0cm}{4ex}$1$}\makebox[0.4cm]{$1$}\makebox[0.4cm]{$1$} \\
   \hline
\end{tabular}
\caption{Truth table of gate F$_3$}
\label{table:GateF3}
\end{center}
\end{table}

Table \ref{table:F3-Operators} shows all the relevant connectives which can be obtained from the
gate. Given the correspondences between the operators of the $3$--valued Zawirski system
$\para{L_3,\oplus}$ and those of the \L ukasiewicz one $\para{L_3,\to_L}$ expressed by equation
\eqref{eq:mv-luk}, we have that the set of connectives realized by the gate F$_3$ satisfies
condition F-4) of universality.

\begin{table}[ht]
\begin{center}
\begin{tabular}{|c|c|c|c|c|}
   \hline
   Connective & Inputs & Constants & Outputs & Garbage \\
   \hline \hline
   \mbox{\rule[0cm]{0cm}{2.5ex}FAN--OUT} & $x_1$ & $x_2 = 1, x_3 = 0$ &
        $y_1, y_2$ & $y_3$ \\
   Pr$_1$   & $x_2, x_3$ & $x_1 = 0$ & $y_3$ & $y_1, y_2$ \\
   Pr$_2$   & $x_2, x_3$ & $x_1 = 0$ & $y_2$ & $y_1, y_3$ \\
   $\oplus$ & $x_1, x_3$ & $x_2 = 1$ & $y_2$ & $y_1, y_3$ \\
   $\odot$  & $x_1, x_2$ & $x_3 = 0$ & $y_2$ & $y_1, y_3$ \\
   Id       & $x_1$ & $x_2 = 0, x_3 = 0$ & $y_1$ & $y_2, y_3$ \\
   $\lnot$  & $x_1$ & $x_2 = 1, x_3 = 0$ & $y_3$ & $y_1, y_2$ \\
   $\sim$   & $x_1$ & $x_2 = 0, x_3 = 1$ & $y_2$ & $y_1, y_3$ \\
   $\pos $  & $x_1$ & $x_2 = 0, x_3 = 1$ & $y_3$ & $y_1, y_2$ \\
   \mbox{\rule[-1.5ex]{0cm}{1.5ex}$\nec$} & $x_3$ & $x_1 = \um, x_2 = 0$ &
        $y_3$ & $y_1, y_2$ \\
   \hline
\end{tabular}
\end{center}
\caption{The operators obtained through gate F$_3$}
\label{table:F3-Operators}
\end{table}

It is worth noting that, as a consequence of Proposition \ref{prop:noFANOUT},
none of the gates presented in this section is strictly conservative.

We conclude this section with the following proposition.

\begin{proposition}
For $d \ge 3$, there is no (3,d)--gate satisfying properties F-2), F-3) and
F-8) which is able to realize the \L ukasiewicz connectives ($\land, \lor,
\to_L$), the G\"odel implication ($\to_G$) and the MV--connectives ($\oplus,
\odot$).
\label{prop:noLMV}
\end{proposition}
\begin{proof}
The only configurations that allow one to realize the classical implication
with a Boolean Fredkin gate are $x_2 = 1$ and $x_3 = 1$. Thus, if we impose
property F-8) to our $(3,d)$--gate we get the following two possibilities to
implement $\to_L$ and $\to_G$:
\begin{center}
\begin{tabular}{|c|c|c|c|c|}
   \hline
   Connective & Inputs & Constant & Output & Garbage \\
   \hline \hline
   \makebox[0.4cm]{\rule[0cm]{0cm}{2.5ex}$\to_L$} & $x_1, x_3$ & $x_2 = 1$ &
        $y_3$ & $y_1, y_2$ \\
   \makebox[0.4cm]{\rule[-1.5ex]{0cm}{2.5ex}$\to_G$} & $x_1, x_2$ & $x_3 = 1$ &
        $y_2$ & $y_1, y_3$ \\
   \hline \hline
   \makebox[0.4cm]{\rule[0cm]{0cm}{2.5ex}$\to_G$} & $x_1, x_3$ & $x_2 = 1$ &
        $y_3$ & $y_1, y_2$ \\
   \makebox[0.4cm]{\rule[-1.5ex]{0cm}{2.5ex}$\to_L$} & $x_1, x_2$ & $x_3 = 1$ &
        $y_2$ & $y_1, y_3$ \\
   \hline
\end{tabular}
\end{center}
However, in both cases there is no configuration that allows one to realize
$\oplus$. In the next table we explore all the configurations and, for each
case, we give a short proof of the incompatibility.
\begin{center}
\begin{tabular}{|c|c|c|c|c|c|}
   \hline
   Connective & Inputs & Constant & Output & Garbage & Proof \\
   \hline \hline
   \makebox[0.4cm]{\rule[0cm]{0cm}{2.5ex}$\oplus$} & $x_1, x_3$ & $x_2 = 1$ &
        $y_2$ & $y_1, y_3$ & (1)\\
   $\oplus$ & $x_1, x_2$ & $x_3 = 1$ & $y_3$ & $y_1, y_2$ & (2)\\
   $\oplus$ & $x_2, x_3$ & $x_1 = \lambda$ & $y_1$ & $y_2, y_3$ & (3)\\
   $\oplus$ & $x_2, x_3$ & $x_1 = \lambda$ & $y_2$ & $y_1, y_3$ & (4)\\
   $\oplus$ & $x_2, x_3$ & $x_1 = \lambda$ & $y_3$ & $y_1, y_2$ & (5)\\
   $\oplus$ & $x_1, x_3$ & $x_2 = \lambda$ & $y_1$ & $y_2, y_3$ & (6)\\
   $\oplus$ & $x_1, x_3$ & $x_2 = \lambda$ & $y_2$ & $y_1, y_3$ & (7)\\
   $\oplus$ & $x_1, x_3$ & $x_2 = \lambda$ & $y_3$ & $y_1, y_2$ & (8)\\
   $\oplus$ & $x_1, x_2$ & $x_3 = \lambda$ & $y_1$ & $y_2, y_3$ & (9)\\
   $\oplus$ & $x_1, x_2$ & $x_3 = \lambda$ & $y_2$ & $y_1, y_3$ & (10)\\
   \makebox[0.4cm]{\rule[-1.5ex]{0cm}{2.5ex}$\oplus$} & $x_1, x_2$ & $x_3 =
        \lambda$ & $y_3$ & $y_1, y_2$ & (11)\\
   \hline
\end{tabular}
\end{center}
where $\lambda \in L_d \setminus \{0,1\}$. In what follows, $\gamma$ is an
unspecified element of $L_d$.

\medskip
\noindent
(1) for $\lambda \leq \um$, all the triples $\lambda 1 (1-\lambda)$ are mapped
to $011$, thus violating F-2);

\noindent
(3) triples $\lambda 1 1$ are mapped to $111$, thus violating F-3);

\noindent
(4) for the triples $\lambda 0 1$, it should be $0 \oplus 1 = 1$ on $y_2$, and
also $\lambda \to_L 0 = 1-\lambda$ or $\lambda \to_G 0= 0$ on the same output;

\noindent
(6) triples $0 \lambda 1$ are mapped to $1 1 \gamma$, thus violating F-3);

\noindent
(7) for the triples $1 \lambda 1$, it should be $1 \oplus 1 = 1$ on $y_2$, and
also $1 \to_L \lambda = 1-\lambda$ or $1 \to_G \lambda = \lambda$ on the same
output;

\noindent
(9) triples $0 1 \lambda$ are mapped to $1 \gamma 1$, thus violating F-3);

\noindent
(10) for the triples $1 1 \lambda$, it should be $1 \oplus 1 = 1$ on $y_2$, and
also $1 \to_L \lambda = 1-\lambda$ or $1 \to_G \lambda = \lambda$ on the same
output;

\noindent
(2),(5),(8),(11) can be obtained respectively from (1), (4), (7), (10) by
exchanging the second and third input/output lines of the gate.
\end{proof}

If a strictly conservative gate that realizes all the three--valued connectives
mentioned above is needed then, due to Propositions \ref{prop:noFANOUT} and
\ref{prop:noLMV}, it is necessary to look for $(n, 3)$--gates with $n \ge 4$.
In a forthcoming paper, a $(4, 3)$--gate which has all the required properties
will be presented.

\section{Finite--valued Universal Gates}

After the discovery of the generalizations of the Fredkin gate to three--valued logics exposed in
the previous section, we obviously tried to generalize further to $d$--valued logics.

The approach followed in the previous section, that is making an exhaustive search in the
space of truth tables of all $(3, d)$--gates, is clearly not feasible to find a solution
which is valid for every value of $d$. As a consequence, we looked for some analytic
expressions which define the new reversible and conservative gates independently of the
cardinality of the set of truth values.

\subsection{A gate for \L ukasiewicz and G\"odel $d$--valued logics}

The first function $\firstfunction: L_d^3 \to L_d^3$ we define is:
\begin{align*}
  & \forall \, \underline{x} = (x_1, x_2, x_3) \in L_d^3 \\
  & \firstfunction (\underline{x}) :=
      \begin{cases}
         (x_1,x_3,x_2) & \text{if $x_1 = 0$ and $x_2\neq x_3$\hspace{2.95cm} i)} \\
         (x_1,x_3,x_2) & \text{if $0 <x_1 \leq x_3 < 1$ and $x_2 = 1$
                         \hspace{1.3cm} ii)} \\
         (x_1,x_3,x_2) & \text{if $0 < x_1 \leq x_2 < 1$ and $x_3 = 1$
                         \hspace{1.2cm} iii)} \\
         (x_1,x_1,1-x_1+x_3) & \text{if $x_3 < x_1 < 1$ and $x_2 = 1$
                               \hspace{1.9cm} iv)} \\
         (x_1,1,x_3+x_1-1) & \text{if $x_1 < 1$, $x_2 = x_1$,
                             $x_3+x_1 \ge 1$} \\
                           & \hspace{0.35cm}\text{and $x_3 < 1$ \hspace{4.15cm}
                             v)} \\
         (x_1,x_1,x_2-x_1) & \text{if $0 < x_1 < x_2 < 1$ and $x_3 = 0$
                             \hspace{1.2cm} vi)} \\
         (x_1,x_3+x_1,0) & \text{if $0 <x_1$, $x_2 = x_1$,
                           $x_3+x_1 < 1$} \\
                         & \hspace{0.35cm}\text{and $x_3 > 0$ \hspace{3.9cm}
                           vii)} \\
         (x_1,x_2,x_3) & \text{otherwise \hspace{4.4cm} viii)}
      \end{cases}
\end{align*}

A direct inspection of the definition allows to conclude that the function
$\firstfunction$ is well defined; that is, each triple $(x_1, x_2, x_3)$ of
$L_d^3$ is associated by $\firstfunction$ with a single triple
$(y_1, y_2, y_3)$ of $L_d^3$.

Let us see some properties of $\firstfunction$.

\begin{proposition}
   $\firstfunction$ is self--reversible.
   \label{pr:f1}
\end{proposition}
\begin{proof}
We have to prove that $\forall \, \underline{x} \in L_d^3$,
$\firstfunction\big(\firstfunction(\underline{x})\big) = \underline{x}$. We
can proceed by dividing the elements of the domain as in rules i), ii), ...,
viii).

Let $a$ and $b$ be two arbitrary elements of $L_d$.
Considering the above rules it holds:

i) $\firstfunction\big(\firstfunction(0,a,b)\big) = \firstfunction(0,b,a) =
(0,a,b)$.

ii) Let $\underline{x} = (a,1,b)$ with $0 < a \leq b < 1$.
Therefore $\underline{y} = \firstfunction(a,1,b) = (a,b,1)$.
$\underline{y}$ fulfills iii) then $\firstfunction(a,b,1) =
(a,1,b) = \underline{x}$.

iii) Let $\underline{x} = (a,b,1)$ with $0 < a \leq b < 1$.
$\underline{y} = \firstfunction(a,b,1) = (a,1,b)$ that fulfills ii)
thus $\firstfunction(a,1,b) = (a,b,1) = \underline{x}$.

iv) Let $\underline{x} = (a,1,b)$ with $b < a < 1$.
$\underline{y} = \firstfunction(a,1,b) = (a,a,1-a+b)$.
Since $0 \leq b < a < 1$, $1-a+b+a \ge 1$ and $1-a+b < 1$.
Therefore $\underline{y}$ fulfills v)
thus $\firstfunction(a,a,1-a+b) = (a,1,1-a+b+a-1) = (a,1,b) = \underline{x}$.

v) Let $\underline{x} = (a,a,b)$ with $a < 1$, $b+a \ge 1$ and $b < 1$.
$\underline{y} = \firstfunction(a,a,b) = (a,1,b+a-1)$.
Since $b+a \ge 1$ and $b < 1$, $0 \leq b+a-1 < a < 1$.
As a consequence we have that $\underline{y}$ satisfies iv),
so $\firstfunction(a,1,b+a-1) = (a,a,1-a+b+a-1) = (a,a,b) = \underline{x}$.

vi) Let $\underline{x} = (a,b,0)$ with $0 < a < b < 1$.
$\underline{y} = \firstfunction(a,b,0) = (a,a,b-a)$.
$b-a+a < 1$ and $b-a > 0$.
This implies that $\underline{y}$ satisfies vii), and consequently
$\firstfunction(a,a,b-a) = (a,b,0) = \underline{x}$.

vii) Let $\underline{x} = (a,a,b)$ with $0 < a$, $b+a < 1$ and $b > 0$.
$\underline{y} = \firstfunction(a,a,b) = (a,b+a,0)$.
We have that $\underline{y}$ satisfies vi), therefore $\firstfunction(a,b+a,0)
= (a,a,b) = \underline{x}$.

viii) obvious.
\end{proof}

The proof of the previous proposition shows the method used to build the
function $\firstfunction$. Rules i) and viii) allow the function to behave as
the Fredkin gate when the inputs are restricted to $\{0,1\}$. Rules ii) and iv)
have been introduced in order to allow the gate to generate the \L ukasiewicz
implication on the third output line and the \L ukasiewicz disjunction on the
second output line when the second input line is set to 1. Rules iii) and v)
are the converse of rules ii) and iv); this is done in order to guarantee the
self--reversibility of the gate. Rules vi) and viii) realize the \L ukasiewicz
conjunction on the second output line when the third input line is set to 0,
whereas vii) and viii) are the converse. Rule viii) uses the simplest
self--reversible function (the identity function) to deal with the cases not
considered by other rules.

Properties F-5), F-6), F-7) and F-8) are trivially satisfied by
$\firstfunction$. Moreover, each rule was written in order to verify the
property of weak conservativeness. In fact, the following proposition holds,
whose proof is straightforward, and thus it is omitted.

\begin{proposition}
$\firstfunction$ is weakly conservative.
\label{pr:f2}
\end{proposition}

It is also easy to see that $\firstfunction$ is a universal function.
In fact, as it is shown in Table \ref{table:f1-Operators}, using suitable
configurations of constants in the input lines we obtain a set of connectives
which suffices to generate, besides the FAN--OUT gate, all the operators of
\L ukasiewicz and G\"odel $d$--valued logics.

\begin{table}[ht]
\begin{center}
\begin{tabular}{|c|c|c|c|c|}
   \hline
   Connective & Inputs & Constants & Outputs & Garbage \\
   \hline \hline
   \mbox{\rule[0cm]{0cm}{2.5ex}FAN--OUT} & $x_1$ & $x_2 = 1, x_3 = 0$ &
        $y_1,y_2$ & $y_3$ \\
   Pr$_1$   & $x_2, x_3$ & $x_1 = 0$ & $y_3$ & $y_1, y_2$ \\
   Pr$_2$   & $x_2, x_3$ & $x_1 = 0$ & $y_2$ & $y_1, y_3$ \\
   $\to_L$  & $x_1, x_3$ & $x_2 = 1$ & $y_3$ & $y_1, y_2$ \\
   $\to_G$  & $x_1, x_2$ & $x_3 = 1$ & $y_2$ & $y_1, y_3$ \\
   $\lor$   & $x_1, x_3$ & $x_2 = 1$ & $y_2$ & $y_1, y_3$ \\
   $\land$  & $x_1, x_2$ & $x_3 = 0$ & $y_2$ & $y_1, y_3$ \\
   Id       & $x_1$ & $x_2 = 0, x_3 = 0$ & $y_1$ & $y_2, y_3$ \\
   $\lnot$  & $x_1$ & $x_2 = 1, x_3 = 0$ & $y_3$ & $y_1, y_2$ \\
   $\sim$   & $x_1$ & $x_2 = 0, x_3 = 1$ & $y_2$ & $y_1, y_3$ \\
   \mbox{\rule[-1.5ex]{0cm}{1.5ex}$\pos$} & $x_1$ & $x_2 =0,
    x_3 = 1$ & $y_3$ & $y_1, y_2$ \\
   \hline
\end{tabular}
\end{center}
\caption{The operators obtained through function $\firstfunction$}
\label{table:f1-Operators}
\end{table}

It is important to underline that, as said before, the properties of the gate
do not depend on the number of truth values involved. Moreover, when $d = 3$
the function $\firstfunction$ behaves just like the gate F$_1$ presented in
the previous section.

\subsection{A family of functions which realize necessity}

Since $\firstfunction$ does not allow one to realize the modal operator $\nec$, we propose the
following family of functions. Let $\lambda$ be any value from the set $L_d \setminus \{0,1\}$; the
family of functions $\secondfunction: L_d^3 \to L_d^3$, parameterized with respect to $\lambda$, is
defined as follows:

\begin{align*}
   & \forall \, \underline{x} = (x_1, x_2, x_3) \in L_d^3 \\
   & \secondfunction (\underline{x}) :=
      \begin{cases}
         (x_2,x_1,x_3) & \text{if $x_1 = 0$, $0 < x_2 < 1$ and
            $x_3 = \lambda$ \hspace{0.95cm} i)} \\
         (x_2,x_1,x_3) & \text{if $0 < x_1 < 1$, $x_2 = 0$ and
            $x_3 = \lambda$ \hspace{0.85cm} ii)} \\
         (x_1,x_3,x_2) & \text{if $x_1 = 0$, $x_2 \neq \lambda$,
                         $x_3 \neq \lambda$ and $x_2 \neq x_3$ iii)} \\
         (x_1,x_3,x_2) & \text{if $0 \leq x_1 \leq x_3 < 1$ and $x_2 = 1$
                         \hspace{1.2cm} iv)} \\
         (x_1,x_3,x_2) & \text{if $0 \leq x_1 \leq x_2 < 1$ and $x_3 = 1$
                         \hspace{1.3cm} v)} \\
         (x_1,x_1,1-x_1+x_3) & \text{if $x_3 < x_1 < 1$ and $x_2 = 1$
                               \hspace{1.9cm} vi)} \\
         (x_1,1,x_3+x_1-1) & \text{if $x_1 < 1$, $x_2 = x_1$,
                             $x_3+x_1 \ge 1$} \\
                           & \hspace{0.35cm}\text{and $x_3 < 1$ \hspace{3.9cm}
                             vii)} \\
         (x_1,x_1,x_2-x_1) & \text{if $0 \leq x_1 < x_2 < 1$ and $x_3 = 0$
                         \hspace{0.95cm} viii)} \\
         (x_1,x_3+x_1,0) & \text{if $0 \leq x_1$, $x_2 = x_1$,
                           $x_3+x_1 < 1$} \\
                         & \hspace{0.35cm}\text{and $x_3 > 0$ \hspace{4cm}
                           ix)} \\
         (x_1,x_2,x_3) & \text{otherwise \hspace{4.7cm} x)}
       \end{cases}
\end{align*}

For each fixed value of $\lambda$ we get a function which realizes the
connectives exposed in Table \ref{table:f2-Operators}.
As it can be seen, the price we pay to realize the modal connective $\nec$
together with all the connectives of $\firstfunction$ is that the functions
$\secondfunction$ loose $0$--regularity in $2d - 5$ input/output pairs and
property F-7) in $2d - 4$ input/output pairs.

Now, let us see some properties of functions $\secondfunction$. The proofs of
the next two propositions are similar to the ones of Propositions~\ref{pr:f1}
and \ref{pr:f2}, and thus they are omitted.

\begin{proposition}
For each fixed value of $\lambda$ in $L_d \setminus \{0,1\}$, the function
$\secondfunction$ is self--reversible.
\end{proposition}

\begin{proposition}
For each fixed value of $\lambda$ in $L_d \setminus \{0,1\}$, the function
$\secondfunction$ is weakly conservative.
\end{proposition}

Properties F-6) and F-8) are trivially satisfied by functions
$\secondfunction$. Table \ref{table:f2-Operators} reports the operators that
can be obtained with the functions $\secondfunction$ by fixing one or two input
lines with constant values from $L_d$. Such configurations have been chosen on
the example of the gate F$_2$ presented in the previous section.

We observe that, also in this case, for a fixed $\lambda$ the constants
involved in such configurations are independent of $d$.

\begin{table}[ht]
\begin{center}
\begin{tabular}{|c|c|c|c|c|}
   \hline
   Connective & Inputs & Constants & Outputs & Garbage \\
   \hline \hline
   \mbox{\rule[0cm]{0cm}{2.5ex}FAN--OUT} & $x_1$ & $x_2 = 1, x_3 = 0$ &
        $y_1, y_2$ & $y_3$ \\
   Pr$_1$   & $x_2, x_3$ & $x_1 = 1$ & $y_2$ & $y_1, y_3$ \\
   Pr$_2$   & $x_2, x_3$ & $x_1 = 1$ & $y_3$ & $y_1, y_2$ \\
   $\to_L$  & $x_1, x_3$ & $x_2 = 1$ & $y_3$ & $y_1, y_2$ \\
   $\to_G$  & $x_1, x_2$ & $x_3 = 1$ & $y_2$ & $y_1, y_3$ \\
   $\lor$   & $x_1, x_3$ & $x_2 = 1$ & $y_2$ & $y_1, y_3$ \\
   $\land$  & $x_1, x_2$ & $x_3 = 0$ & $y_2$ & $y_1, y_3$ \\
   Id       & $x_1$ & $x_2 = 0, x_3 = 0$ & $y_1$ & $y_2, y_3$ \\
   $\lnot$  & $x_1$ & $x_2 = 1, x_3 = 0$ & $y_3$ & $y_1, y_2$ \\
   $\sim$   & $x_1$ & $x_2 = 0, x_3 = 1$ & $y_2$ & $y_1, y_3$ \\
   $\pos$   & $x_1$ & $x_2 = 0, x_3 = 1$ & $y_3$ & $y_1, y_2$ \\
   \mbox{\rule[-1.5ex]{0cm}{1.5ex}$\nec$} & $x_1$ & $x_2 = 0, x_3 = \lambda$
        & $y_1$ & $y_2, y_3$ \\
   \hline
\end{tabular}
\end{center}
\caption{The operators obtained through functions $\secondfunction$}
\label{table:f2-Operators}
\end{table}

Note that, when $d = 3$, the function $\underline{f}_{d,\um}^2$ behaves just
like the gate F$_2$ presented in the previous section.

\subsection{A gate for MV--connectives}

None of the gates just presented generates the MV connectives showed in Section
\ref{section:MVLogic}. This fact led us to build the function $\mv : L_d^3 \to L_d^3$ defined as
follows:

\begin{align*}
   & \forall \, \underline{x} = (x_1,x_2,x_3) \in L_d^3 \\
   & \mv (\underline{x}) :=
      \begin{cases}
         (x_1,x_3,x_2) & \text{if $x_1 = 0$ and $x_2 \neq x_3$ \hspace{2.7cm} i)} \\
         (x_1,x_1+x_3,1-x_1) & \text{if $x_1 > 0$, $x_2 = 1$ and
                               $x_1 + x_3 < 1$ \hspace{0.6cm} ii)} \\
         (x_1,1,x_2-x_1) & \text{if $0 < x_1 \leq x_2 < 1$ and $x_3 = 1-x_1$
                           \hspace{0.2cm} iii)} \\
         (x_1,x_1+x_2-1, & \\
         \hspace{2.15cm}1-x_1) & \text{if $x_1 < 1$, $x_2 < 1$, $x_3 = 0$
                                 and} \\
                               & \hspace{0.35cm} \text{$x_1+x_2 > 1$
                                 \hspace{3.8cm} iv)} \\
         (x_1,x_2+x_3,0) & \text{if $0 < x_2 < x_1 < 1$ and $x_3 = 1 - x_1$
                             \hspace{0.35cm} v)} \\
         (x_1,x_3,x_2) & \text{if $0 < x_1$, $x_2 > 0$, $x_3 = 0$ and} \\
                       & \hspace{0.35cm} \text{$x_1 + x_2 \leq 1$
                         \hspace{3.8cm} vi)} \\
         (x_1,x_3,x_2) & \text{if $0 < x_1$, $x_2 = 0$, $x_3 > 0$ and} \\
                       & \hspace{0.35cm} \text{$x_1 + x_3 \leq 1$
                         \hspace{3.7cm} vii)} \\
         (x_1,x_2,x_3) & \text{otherwise \hspace{4.3cm} viii)}
      \end{cases}
\end{align*}

In order to find this gate we used the technique previously shown: first we
looked at the gate F$_3$ exposed in the previous section in order to know which
configurations give rise to the operators $\oplus$ and $\odot$; successively,
we wrote their inverses. Thus it is no wonder that, for $d = 3$, the function
$\mv$ behaves like the gate F$_3$ presented in the previous section.

As for the previous functions, we can state the following properties.

\begin{proposition}
$\mv$ is self--reversible.
\end{proposition}

\begin{proposition}
$\mv$ is weakly conservative.
\end{proposition}

Moreover, properties F-5), F-6), F-7) and F-8) are trivially satisfied by
$\mv$.

Table \ref{table:mv-Operators} reports the operators that can be obtained from function $\mv$ by
fixing one or two input lines with constant values from $L_d$. As we can see, $\mv$ is a gate
providing functional completeness of finite--valued calculus, regardless of the value assumed by
$d$.

\begin{table}[ht]
\begin{center}
\begin{tabular}{|c|c|c|c|c|}
   \hline
   Connectives & Inputs & Constants & Outputs & Garbage \\
   \hline \hline
   \mbox{\rule[0cm]{0cm}{2.5ex}FAN--OUT} & $x_1$ & $x_2 = 1, x_3 = 0$ &
        $y_1, y_2$ & $y_3$ \\
   Pr$_1$   & $x_2, x_3$ & $x_1 = 0$ & $y_3$ & $y_1, y_2$ \\
   Pr$_2$   & $x_2, x_3$ & $x_1 = 0$ & $y_2$ & $y_1, y_3$ \\
   $\oplus$ & $x_1, x_3$ & $x_2 = 1$ & $y_2$ & $y_1, y_3$ \\
   $\odot$  & $x_1, x_2$ & $x_3 = 0$ & $y_2$ & $y_1, y_3$ \\
   Id       & $x_1$ & $x_2 = 0, x_3 = 0$ & $y_1$ & $y_2, y_3$ \\
   $\lnot$  & $x_1$ & $x_2 = 1, x_3 = 0$ & $y_3$ & $y_1, y_2$ \\
   $\sim$   & $x_1$ & $x_2 = 0, x_3 = 1$ & $y_2$ & $y_1, y_3$ \\
   $\pos$   & $x_1$ & $x_2 = 0, x_3 = 1$ & $y_3$ & $y_1, y_2$ \\
   \mbox{\rule[-1.5ex]{0cm}{1.5ex}$\nec$} & $x_3$ & $x_1 = \frac{1}{d-1}$,
        $x_2 = 0$ & $y_3$ & $y_1, y_2$ \\
   \hline
\end{tabular}
\end{center}
\caption{The operators obtained through function $\mv$}
\label{table:mv-Operators}
\end{table}

\section{Conclusions and Directions for Future Work}

We presented some generalizations of the Fredkin gate for $d$--valued reversible and conservative
logics, notably $d$--valued \L ukasiewicz and $d$--valued G\"odel logics. In particular, we
introduced three gates for three--valued logics and three possible extensions of such gates for
$d$--valued logics; one of the extensions was specifically designed to realize the
MV--connectives. Moreover we showed how to realize, with such gates, the operators that
characterize some modal logics.

One of the purposes of our work was to show that the framework of reversible
and conservative computation can be extended toward some non classical
``reasoning environments'', originally proposed to deal with propositions
which embed imprecise and uncertain information, that are usually based upon
many--valued and modal logics.

It remains open the question on how it is possible to extend further the
framework towards infinite--valued logics, such as fuzzy logics, both with
$\aleph_0$ and $\aleph_1$ truth values. We feel that in such settings many new
and interesting questions arise; here we propose just a few of them. For
example: since reversible circuits need no more to have the same number of
input and output lines, and moreover we can encode on a single input (or
output) as much information as we want, what are the computational properties
of such circuits? What are the differences with respect to reversible and
conservative circuits for $d$--valued logics? How can we characterize the set
of functions computed by such circuits?

Moreover, it is not difficult to extend Proposition \ref{prop:noFANOUT} to deal with an
infinite number of truth values. A direct consequence is that there are no possible
extensions of the Fredkin gate to infinite--valued logics which compute the FAN--OUT
function and at the same time are strictly conservative. How does this change the notion
of conservativeness? In this paper we proposed the alternative notion of weak
conservativeness, together with a possible physical interpretation; however, when dealing
with an infinite number of energy levels there are two possibilities: either the energy
levels extend over an unlimited range, so that to switch from a given level to another it
could be necessary an infinite amount of energy, or the levels become increasingly close
to each other. In the latter case, an infinite precision on the amount of energy can be
required to switch from one level to another; in this situation, when the energy gap
between the levels becomes smaller than the underlying thermal noise the computing
physical system goes out of control. The above observations lead naturally to the
following question: are the circuits for infinite--valued logics physically realizable?
On the other hand, do we really need them?

\newpage
\section{Appendix: The abstract algebraic approaches to many--valued logics}

\subsection{$BZW$ algebras}
The set of numbers $L_\alpha$ ($\alpha\in\{d,\aleph_0,\aleph_1\}$), interpreted as possible truth
values of propositional sentences, equipped with the connectives $\to_L$, $\neg$, and $\sim$ are
standard models of an abstract system, called Brouwer--Zadeh--Wajsberg ($BZW$) algebra, which can
be considered a useful algebraic environment of many--valued logics.

Wajsberg ($W$) algebras were introduced by Wajsberg in order to give an algebraic axiomatization to
many valued logics \cite{Wa31, Wa35}. Fundamental aspect of $W$ algebras is the usage of \L
ukasiewicz implication as a primitive operator. Brouwer--Zadeh ($BZ$) lattices, on the other hand,
involve an intuitionistic  negation $\sim$, besides a fuzzy one $\lnot$ \cite{cattaneo-nistico}.
Moreover, by suitable compositions of the two negations, it is possible to define the two basic
modal operators, {\em necessity} ($\nec$) and {\em possibility} ($\pos$). Thus, by a pasting of the
two structures one obtains $BZW$ algebras. These algebras also result to be a general classical
``unsharp environment'' for an abstract introduction to rough approximation spaces.
\begin{definition}
 \label{df:bzw}
   A Brouwer Zadeh Wajsberg ($BZW$) algebra is a system $\langle A, \rightarrow, \lnot, \sim, 1 \rangle$,
   where $A$ is a nonempty set, $1$ is a constant element, $\lnot$ and $\sim$ are unary operators, and
   $\rightarrow$ is a binary operator, obeying the following axioms:
\begin{description}
    \item[($BZW1$)]  $1 \rightarrow x = x$
    \item[($BZW2$)] $(x \rightarrow y) \rightarrow ((y \rightarrow z) \rightarrow (x \rightarrow z))=1$
    \item[($BZW3$)] $(x \rightarrow y) \rightarrow y = (y \rightarrow x) \rightarrow x$
    \item[($BZW4$)] $(\lnot x \rightarrow \lnot y) \rightarrow (y \rightarrow x)=1$
    \item[($BZW5$)] $\lnot \sim x \rightarrow \sim\sim x =1$
    \item[($BZW6$)] $(\lnot x \rightarrow \sim\sim x )\rightarrow \sim\sim x = 1$
    \item[($BZW7$)] $\lnot \sim ((x \rightarrow y) \rightarrow y)=
         (\lnot\sim x \rightarrow \sim\sim y) \rightarrow \sim\sim y$
\end{description}
A {\it de Morgan} $BZW$, shortly a $BZW^{dM}$, algebra is a $BZW$ algebra in which axiom ($BZW7$)
is replaced by the
  following:
   \begin{description}
    \item[$(BZW7')$] $\sim\lnot[(\lnot x \rightarrow \lnot y)\rightarrow \lnot y] =
    (\lnot \sim\sim x\rightarrow \lnot \sim\sim y)\rightarrow \lnot \sim\sim y$
  \end{description}
\end{definition}

Let us note that the substructure $\langle A,\rightarrow,\lnot, 1\rangle $ induced by a $BZW$ algebra is a $W$ algebra,
just characterized by axioms ($BZW1$)--($BZW4$)~\cite{Wa31, Wa35}.
Now, from any $BZW$ algebra it is possible to obtain in a canonical way a structure of {\em BZ
distributive lattice} according to the following result.
\begin{theorem}
\label{th:bzw2bz}
   Let $\langle A,\rightarrow,\lnot,\sim, 1 \rangle$ be a $BZW$ algebra.
   Let us introduce a new constant and two derived operators according to the following:
\begin{align}
  0 &:= \lnot 1 \label{eq:def1}\\
  x \lor y &:= (x \rightarrow y) \rightarrow y\label{eq:def4}\\
  x \land y &:= \lnot ( (\lnot x \rightarrow \lnot y) \rightarrow \lnot y)\label{eq:def5}
\end{align}
   Then the structure $\para {A,\land,\lor,\lnot,\sim, 0}$ is a distributive $BZ$ lattice. In other words:
  \begin{enumerate}
\item[(1)]
  $A$ is a  distributive lattice with respect to  the join and the meet
  operations $\lor,\land $ defined by (\ref{eq:def4}) and (\ref{eq:def5}), respectively. The partial
  order relation induced by these operations is: 
   \begin{equation}
   \label{eq:order}
     x\leq y \quad \text{iff}\quad  x\rightarrow y=1.
   \end{equation}
   $A$ is bounded by the least element $0$ and the
   greatest element $1$: $$ \forall x\in A, \quad 0\le x\le 1. $$
\item[(2)]
The unary operation $\lnot:A\mapsto A$ is a {\rm Kleene} (or {\rm Zadeh}) orthocomplementation. In
other words the following hold:
\begin{enumerate}\item[]
  \begin{enumerate}
\item[(K1)] \quad
        $\lnot (\lnot x)=x$
\item[(K2)] \quad
        $\lnot (x\lor y)=\lnot x\land \lnot y $
\item[(K3)]\quad
        $x\wedge \lnot x \le y\vee \lnot y.$
\end{enumerate}
   \end{enumerate}
\item[(3)]
The unary operation $\sim:A\mapsto A$ is a {\rm Brouwer} orthocomplementation. That is, it
satisfies the following properties:
\begin{enumerate}\item[]
\begin{enumerate}
\item[(B1)] \quad
        $x\wedge \sim \sim x= x $
\item[(B2)]\quad
        $\sim (x\vee y)=\sim x\wedge \sim y $
\item[(B3)]\quad
        $x\wedge \sim x=0 $
\end{enumerate}
\end{enumerate}
\item[(4)]
The two orthocomplementations are linked by the following interconnection rule:
\begin{enumerate}\item[]
\begin{enumerate}
\item[(in)]\quad $\lnot \sim x=\sim \sim x $
 \end{enumerate}
\end{enumerate}
\end{enumerate}
\end{theorem}

    Let us note that under condition ($K1$) the {\em de Morgan} law ($K2$)
is equivalent to the dual de Morgan law ``$\lnot (x \land y) = \lnot x \land \lnot y$'' and to the
Kleene {\em contraposition} law ``$x \leq y$ implies $\lnot y\leq \lnot x$''. In general neither
the {\em noncontradiction} law ``$\forall x: x\land \lnot x = 0$'' nor the {\em excluded--middle}
law ``$\forall x: x\lor \lnot x = 1$'' hold from this negation, even if for some elements $e$ (for
instance $e =0,1$) it may happen that $e \land \lnot e=0$ and $e\lor \lnot e =1$.

As to the Brouwer negation, we have that under condition ($B1$) the de Morgan law ($B2$) is
equivalent to the Brouwer contraposition law ``$x\leq y$ implies $\sim y \leq \sim x$'', but not to
the dual de Morgan law. In this case the intuitionistic {\em noncontradiction} law is verified, but
the excluded middle law in general is not required to hold.

On the other hand, if the structure in Theorem \ref{th:bzw2bz} is a $BZW^{dM}$ algebra then the
Brouwer negation satisfies also the dual de Morgan law ``$\sim (x \land y)= \sim x \lor \sim y$''.
However, in both cases (either $BZW$ or $BZW^{dM}$) in general the Brouwer negation satisfies the
{\em weak double negation} law ($B1$), also written as ``$\forall x: x \leq \sim\sim x$'', which
does not forbid that for some special elements $e$ (for instance $e=0,1$) $e = \sim\sim e$ holds.

A third kind of complementation, called {\it anti--intuitionistic orthocomplementation\/}, can be
defined in any $BZW$ algebra.
\begin{definition}
   Let ${\cA}=\para{A, \rightarrow, \lnot, \sim, 1}$ be a $BZW$ algebra.
   The {\it anti--in\-tui\-tio\-nis\-tic complementation} is the unary
   operation $\flat: A\mapsto A$ defined as follows:
   $$
    \flat x:=\lnot \sim \lnot x
   $$
\end{definition}
One can easily show that $\flat$ satisfies the following conditions:
\begin{enumerate}\item[]
   \begin{enumerate}
    \item[(AB1)]\quad
    $\flat \flat x\leq x$;
    \item[(AB2)]\quad
    $\flat x\lor \flat y=\flat (x\land y)$\quad[equivalently, $x\leq
    y$ implies $\flat y\leq \flat x$];
    \item[(AB3)]\quad
    $x\lor \flat x=1$.
\end{enumerate}
\end{enumerate}

As we have said at the beginning of this Appendix, the structure
 \break
$\para{L_\alpha,\,\to_L,\,\neg,\,\sim,1}$ based on the set of truth values $L_\alpha$ from the real
unit interval, is a model of $BZW^{dM}$ algebraic structure with respect to the \L ukasiewicz
implication connective $\to_L$, the diametrical negation $\neg$, and the impossibility negation
$\sim$ introduced in Section~\ref{section:MVLogic}.
\subsection{Modal operators in BZW algebras}
Modal operators can be naturally introduced in any Brouwer Zadeh lattice (hence in any $BZW$
algebra). The {\em necessity operator\/} $\nec$ and the {\em possibility operator\/} $\pos$ are
defined in terms of Zadeh and Brouwer complementations.
\begin{definition}
   For any element $x$ of a Brouwer
   Zadeh lattice ${\cA}$, the {\it necessity\/} and the {\it
   possibility\/} of $x$ are defined as follows:
\\
   (n)\quad The {\em necessity\/}:
    $
    \nec(x):=\sim\neg x
    $.
\\
   (p)\quad The {\em possibility\/}:
    $
    \pos(x):=\neg\nec(\neg x)
    $.
\end{definition}

As a consequence, one obtains:
\begin{align*}
\pos(x)=\neg \sim x \qquad &\nec (\pos (x))=\sim \sim x
\\
\sim x=\nec (\neg x)=\neg(\pos x) \qquad &\flat x=\neg(\nec x)=\pos(\neg x)
\end{align*}

On this basis, similarly to the modal interpretation of intuitionistic logic, the Brouwer
complementation $\sim$ can be interpreted as the negation of possibility or {\em impossibility}
(also the necessity of a negation). Analogously, the anti--Brouwer complementation $\flat$ can be
interpreted as the negation of necessity or {\em contingency\/}.

Our modal operators $\nec$ and $\pos$ turn out to have an $S_5$--like behavior based on a Kleene
algebra, rather than on a Boolean one. Since $\neg$ represents here a {\em fuzzy} (i.e., {\em
Kleene}) negation on a distributive lattice, the result will be a {\em fuzzy} (i.e., {\em Kleene})
$S_5$ modal situation.

\begin{theorem}
\label{th:mod} In any BZ lattice the following conditions hold:
\begin{enumerate}
\item[(1)]
                $$\nec(x)\le x \le \pos(x)$$
   In other words: necessity implies actuality and actuality implies
   possibility (a characteristic principle of the modal system $T$ \cite{chellas}).
\item[(2)]
   $$\nec(\nec(x))=\nec(x)$$ $$\pos(\pos(x))=\pos(x)$$ Necessity of
   necessity is equal to necessity; similarly for possibility (a
   characteristic $S_4$--principle \cite{chellas}).
\item[(3)]
   $$x\le \nec(\pos(x))$$ Actuality implies necessity of possibility (a
   characteristic $B$--principle \cite{chellas}).
\item[(4)]
   $$\pos(x)=\nec(\pos(x))$$ $$\nec(x)=\pos(\nec(x))$$ Possibility is equal
   to the necessity of possibility; analogously, necessity is equal to the
   possibility of necessity (a characteristic $S_5$--principle \cite{chellas}).
\end{enumerate}
\end{theorem}

On this basis, the definition of $BZW$ algebras admits of a natural modal translation. It is
worthwhile noticing that the modal translation of axiom ($BZW6$) ($\neg x \rightarrow
\nec(\pos(x))=\nec(\pos(x))$ ) asserts a weak (modal) version of the {\it consecutio mirabilis\/}
principle. As it is well known, the strong {\it consecutio mirabilis\/} principle ($((\neg
x\rightarrow x) \rightarrow x)=1$ ) is not generally valid in the case of $BZW$ algebras.

\subsection{Rough approximation spaces in BZW algebras}
As stated in Theorem \ref{th:mod}, in general the order chain $\nec(x)\le x \le \pos(x)$ holds.
Clearly, this is a fuzzy situation. In a crisp environment we have no difference among necessity,
actuality and possibility, i.e., we are interested to those elements for which $ e =\pos(e)$
(equivalently, $ e =\nec(e)$). This leads one to define the substructure of all {\it M--sharp}
(exact, crisp) elements, denoted by $A_{e,M}$, as follows:
\begin{align*}
    A_{e,M}:= \{e \in A: \pos(e)=e\}
              = \{e\in A: \nec(e)=e\}
\end{align*}
However, this is not the only way to define sharp elements. In fact, since in general $x\land \lnot x \neq 0$
(equivalently, $x\lor \lnot x \neq 1$)
 it is possible to consider
as {\it Kleene sharp} (K--sharp)  the elements which satisfy the non contradiction (or,
equivalently, the excluded middle) law with respect to the Kleene negation:
\begin{align*}
    A_{e,\lnot}:= \{e \in A: e\land \lnot e=0\}  = \{e \in A: e\lor \lnot e=1\}
\end{align*}
Alternatively, considering the Brouwer negation we have that the {\it weak double negation} law
holds ($\forall x \in A, x\leq \sim\sim x$) whereas the  double negation law fails. So we can
introduce a further definition of {\it Brouwer sharp} (B--sharp) elements as follows:
\begin{align*}
    A_{e,\sim}:= \{e \in A: \sim\sim e=e\} = \{e\in A: \flat\flat e=e\}
\end{align*}
Finally, as stated before the property $\lnot x\rightarrow (\nec(\pos(x))=(\nec(\pos(x)))$ holds
but in general it is not true that $\lnot x\rightarrow x=x$. As a consequence, the
$\rightarrow$--sharp elements are:
$$
    A_{e,\rightarrow}:= \{e \in A: \lnot e\rightarrow e=e\}
$$
The relation among all these different substructures of exact elements is figured out in the
following proposition.
\begin{proposition}
   Let ${\it A}$ be a $BZW$ algebra. Then
$$
    A_{e,\sim}=A_{e,M}\subseteq  A_{e,\rightarrow}=A_{e,\lnot}
$$
Let ${\it A}$ be a $BZW^{dM}$ algebra. Then
$$
    A_{e,\sim}=A_{e,M}=  A_{e,\rightarrow}=A_{e,\lnot}
$$
\end{proposition}
Consequently, in the case of $BZW^{dM}$ algebras we simply talk of {\em sharp} elements and write
$A_e$. Otherwise, in the more general case of $BZW$ algebras we distinguish between {\em B--sharp}
elements, i.e., elements in $A_{e,\sim}$ ($=A_{e,M}$), and {\em K--sharp} elements, i.e., elements
belonging to $A_{e,\lnot}$ ($=A_{e,\rightarrow}$).

As we have seen, in any $BZW$ algebra it is possible, through the composition of  the two
negations, to introduce the modal operators $\nec$ and $\pos$. These operators can be used to give
a rough approximation of any element $x \in A$ by B--sharp definable elements. In fact, $\nec(x)$
(resp., $\pos(x)$) turns out to be  the best approximation from the bottom (resp., top) of $x$ by
B--sharp elements. To be precise, for any element $x \in A$ the following holds:
\begin{enumerate}
\item[]
\begin{enumerate}
   \item[(I1)] $\nec(x)$ is B--sharp ($\nec (x) \in A_{e,\sim}$)
   \item[(I2)] $\nec(x)$ is an {\em inner} ({\em lower}) approximation of $x$ ($\nec (x) \leq x$)
   \item[(I3)] $\nec(x)$ is the best inner approximation of $x$ by B--sharp elements
     (let $e \in  A_{e,\sim}$ be
    such that $e\leq x$, then $e\leq \nec(x)$)
\end{enumerate}
\end{enumerate}
Analogously,
\begin{enumerate}
\item[]
\begin{enumerate}
   \item[(O1)] $\pos(x)$ is B--sharp ($\pos (x) \in A_{e,\sim}$)
   \item[(O2)] $\pos(x)$ is an {\em outer} ({\em upper}) approximation of $x$ ($x \leq \pos (x)$)
   \item[(O3)] $\pos(x)$ is the best outer approximation of $x$ by B--sharp elements
     (let $f \in  A_{e,\sim}$ be such that $x\leq f$, then $\pos(x)\leq f$)
\end{enumerate}
\end{enumerate}
\begin{definition}
   Given a $BZW$ algebra $\langle A, \rightarrow, \lnot, \sim, 1 \rangle$,
   the induced {\em rough approximation space} is the structure
   $\langle A,A_{e,\sim},\nec,\pos\rangle$ consisting of the set $A$ of all the elements which can be
   approximated, the set  $A_{e,\sim}$ of all {\em definable} (or B--sharp) elements, and the {\em inner}
   (resp., {\em outer}) {\em approximation map} $\nec: A \to A_{e,\sim}$ (resp., $\pos: A \to A_{e,\sim}$).

For any element $x \in A$, its {\em rough approximation} is defined as the pair of B--sharp
elements:
   $$
    r(x) := \langle \nec(x),\pos(x)\rangle \quad
        [\text{with}\quad \nec(x) \leq x \leq \pos(x)]
   $$
   drawn in the following diagram:
$$
\xymatrix{
  &
  x\in A \ar[dl]_{\nec}\ar[dr]^{\pos} \ar[dd]^r
  &
  \\
  \nec(x)\in A_{e,\sim} \ar[dr]
  &&
  \pos(x) \in A_{e,\sim}\ar[dl]
  \\
  &
  \langle\nec(x),\pos(x)\rangle
  &
     }
$$
\end{definition}
So the mapping $r: A\to A_{e,\sim} \times A_{e,\sim}$ approximates an unsharp (fuzzy) element by a
pair of B--sharp (crisp, exact) ones representing its inner and outer sharp approximation,
respectively. Clearly, B--sharp elements are characterized by the property that they coincide with
their rough approximations:
$$
   e\in A_{e,\sim} \quad \text{iff} \quad r(e)= \langle e,e \rangle.
$$
\subsection{$BZW$ algebras and $BZMV$ algebras}
In Section \ref{section:MVLogic} we have seen that in each $L_\alpha$ the identities $x\to_L y=\neg
x\oplus y$ and $x\oplus y =\neg x\to_L y$ show that there is no essential difference of expressive
power between the implication connective $\to_L$ and the additive one $\oplus$, owing to their
mutual interdefinability. This suggests to introduce an algebraic structure, called $BZMV$ algebra
(see \cite{CDG98}, \cite{CGP99}), based on the primitive connective $\oplus$, and to show that
$BZMV$ algebras and  $BZW$ algebras are categorically equivalent. First of all we recall the
definition of $BZMV$ algebras.
\begin{definition}
   A $BZMV$ algebra is a system $\langle A, \oplus, \lnot, \sim, 0 \rangle$, where $A$ is a nonempty set,
   0 is  a constant, $\lnot$ and $\sim$ are unary operators, and $\oplus$ is a binary operator, obeying the
   following axioms:
\begin{description}
    \item[($BZMV1$)] $ (x \oplus y) \oplus z  = (y \oplus z) \oplus x$
    \item[($BZMV2$)] $x \oplus 0=x$
    \item[($BZMV3$)] $\lnot (\lnot x) = x$
    \item[($BZMV4$)] $\lnot(\lnot x \oplus y) \oplus y = \lnot (x \oplus \lnot y) \oplus x$
    \item[($BZMV5$)] $ \sim x \oplus \sim\sim x = \lnot 0$
    \item[($BZMV6$)] $ x \oplus \sim\sim x = \sim\sim x$
    \item[($BZMV7$)] $\lnot \sim [(\lnot(\lnot x \oplus y) \oplus y)]=
        \lnot(\sim x \oplus \sim\sim y) \oplus \sim\sim y$
\end{description}
\end{definition}

Now we introduce in a $BZW$ algebra two new operators:
\begin{align}
  x \oplus y &:= \lnot x \rightarrow y \label{eq:def2}\\
  x \odot y &:= \lnot( x \rightarrow \lnot y)\label{eq:def3}
\end{align}
As to the relationship between the two structures of $BZW$ and $BZMV$ algebras we have the
following theorem.
\begin{theorem}
\par\noindent
\begin{enumerate}
\item
   Let $\cA=\langle A, \rightarrow, \lnot, \sim, 1 \rangle$ be a $BZW$ algebra. Then putting
   $x \oplus y:= \lnot x\rightarrow y$ and $0:= \lnot 1$, the corresponding system
   $\cA_{BZMV}=\langle A, \oplus, \lnot, \sim, 0 \rangle$ is a $BZMV$ algebra.
\item
   Let $\cA=\langle A, \oplus, \lnot, \sim, 0 \rangle$ be a $BZMV$ algebra. Then putting
   $x \rightarrow y:= \lnot x\oplus y$ and $1:= \lnot 0$, the corresponding system
   $\cA_{BZW}=\langle A, \rightarrow, \lnot, \sim, 1 \rangle$ is a $BZW$ algebra.
\item
\begin{enumerate}
\item
Let $\cA$ be any $BZW$ algebra, then
    $
       \cA=(\cA_{BZMV})_{BZW}
    $
\item Let  $\cA$ be any $BZMV$ algebra, then
    $
        \cA=(\cA_{BZW})_{BZMV}
    $
\end{enumerate}
\end{enumerate}
\end{theorem}

Thus we have that under a suitable definition of the involved operators, $BZW$ and $BZMV$ algebras
are categorically equivalent. Being no difference between the expressive power of the two
structures, one can choose the algebra that best fits his/her analysis.

In particular, in the context of $BZMV$ algebras the partial order relation defined by
\eqref{eq:order}, taking into account \eqref{eq:def2}, assumes the form:
\begin{equation}
\label{eq:mv-ord}
  x\le y\quad\text{iff}\quad \neg x\oplus y = 1
\end{equation}
Moreover, the induced structure of {\em $BZ$ distributive lattice} $\para{A,\lor,\land,\neg\sim,0}$
of Theorem~\ref{th:bzw2bz} in the present case is obtained from the  operations of meet and join,
intrinsically defined using the operators $\oplus$ and $\lnot$ as follows:
\begin{align*}
   & x \lor y := (x \odot \lnot y) \oplus y = \lnot (\lnot x \oplus y) \oplus y \\
   & x \land y := (x \oplus \lnot y) \odot y = \lnot [\lnot (x \oplus \lnot y)
                  \oplus \lnot y]
\end{align*}

As to the Kleene sharp elements  we have now the following identifications:
\begin{align*}
A_{e,\neg} =\{e\in A:e\oplus e=e\} =\{e\in A:e\odot e=e\}
\end{align*}

In the case of $BZMV$ algebras, the set $A_{e,\neg}$ of all Kleene sharp elements is closed under
the operations $\oplus$, $\odot$, $\neg$, and $\sim$; moreover on this set the lattice operations
$\lor$ and $\land$ coincide with the $MV$ algebra operations $\oplus$ and $\odot$ (see \cite{CDG98,
CGP99}) :
$$
\forall e,f\in A_{e,\neg},\quad e\oplus f=e\lor f\quad\text{and}\quad e\odot f=e\land f
$$
The structure $\para{A_{e,\neg},\oplus,\odot,\neg,\sim,0}$ is the largest $BZMV$ subalgebra of $A$
which is at the same time a Boolean $BZ$ lattice with respect to the operations $\lor(=\oplus)$,
$\land(=\odot)$, $\neg$ and $\sim$. That is, $A_{e,\neg}$ is a $BZ$ distributive lattice such that
the substructure $\para{A_{e,\neg},\lor,\land,\neg,0}$ is a Boolean (rather than just a Kleene)
lattice.

Relatively to the Brouwer sharp elements, the set $A_{e,\sim}$ is closed under the operations
$\oplus$, $\odot$, $\neg$, and $\sim$. Moreover,
\begin{gather*}
\forall e,f\in A_{e,\sim},\quad e\oplus f=e\lor f\quad\text{and}\quad e\odot f=e\land f
\\
\forall e\in A_{e,\sim},\quad \neg e=\sim e
\end{gather*}
Analogously, the structure $\para{A_{e,\neg},\oplus,\neg,0}$ is the largest $MV$ subalgebra of $A$
which is at the same time a Boolean lattice with respect to the same operations $\lor(=\oplus)$,
$\land(=\odot)$, and $\neg$.
\subsection{Chang and Wajsberg many--valued algebras}
MV--algebras are algebraic structures introduced by C.C.~Chang in order to provide an algebraic
proof of the completeness theorem for the infinite many--valued logic of \L ukasiewicz (see
\cite{chang-58} and \cite{chang-59}). A privileged model of this logic is based on the set
$L_{\aleph_1}$ of truth values, which gives rise to a totally ordered MV--algebra. Here we present
a definition of MV--algebra which is simpler than the axiomatization proposed by Mangani in
\cite{mangani}.

\begin{definition}
An MV--algebra is a structure $\parang{L,\oplus,\lnot,0}$ where $L$ is a nonempty set, $0$ is a
constant element of $L$, $\oplus$ is a binary operation on $L$, and $\lnot$ is a unary operator on
$L$, satisfying the following axioms:
\begin{lquote}
\begin{itemize}
   \item[P1)] $(x \oplus y) \oplus z = (y \oplus z) \oplus x$
   \item[P2)] $x \oplus 0 = x$
   \item[P3)] $x \oplus \lnot 0 = \lnot 0$
   \item[P4)] $\lnot (\lnot 0) = 0$
   \item[P5)] $\lnot (\lnot x \oplus y) \oplus y = \lnot (x \oplus \lnot y)
               \oplus x$
\end{itemize}
\end{lquote}
\end{definition}
Axioms P1) -- P5) are independent, as it is shown in \cite{cattaneo-lombardo}.
In \cite{CDG98} the following result is proved.
\begin{proposition}
Let $\para{A,\oplus,\neg,\sim,0}$ be a $BZMV$ algebra. Then the substructure
$\para{A,\oplus,\neg,0}$ is an $MV$ algebra.
\end{proposition}

Using $\oplus$ and $\lnot$, in this algebraic context we can define the derived operations:
\begin{align*}
   & 1 := \lnot 0 \\
   & x \odot y := \lnot (\lnot x \oplus \lnot y) \\
   & x \lor y := (x \odot \lnot y) \oplus y = \lnot (\lnot x \oplus y) \oplus y \\
   & x \land y := (x \oplus \lnot y) \odot y = \lnot [\lnot (x \oplus \lnot y)
                  \oplus \lnot y]
\end{align*}
obtaining a structure $\parang{L, \oplus, \odot, \lor, \land, \lnot, 0, 1}$ which satisfies the
following conditions, assumed as axioms by Chang in his original definition:

\begin{center}
   \begin{tabular}{llcll}
        (C1)  & $x \oplus y = y \oplus x$ &\qquad
      & (C1') & $x \odot y = y \odot x$             \\
        (C2)  & $x \oplus (y \oplus z) = (x \oplus y) \oplus z$ &{}
      & (C2') & $x \odot (y \odot z) = (x \odot y) \odot z$ \\
        (C3)  & $x \oplus \lnot x = 1$ &{}
      & (C3') & $x \odot \lnot x = 0$           \\
        (C4)  & $x \oplus 1 = 1$ & {}
      & (C4') & $x \odot 0 = 0$ \\
        (C5)  & $x \oplus 0 = x$ & {}
      & (C5') & $x \odot 1 = x$ \\
        (C6)  & $\lnot (x \oplus y) = \lnot x \odot \lnot y$ & {}
      & (C6') & $\lnot (x \odot y) = \lnot x \oplus \lnot y$ \\
        (C7)  & $\lnot (\lnot x) = x$ & {}
      & (C8)  & $\lnot 0 = 1$ \\          
        (C9)  & $x \lor y = y \lor x$ & {}
      & (C9') & $x \land y = y \land x$ \\
        (C10) & $x \lor (y \lor z) = (x \lor y) \lor z$ & {}
      & (C10')& $x \land (y \land z) = (x \land y) \land z$   \\
        (C11) & $x \oplus (y \land z) = (x \oplus y) \land (x \oplus z)$ & {}
      & (C11')& $x \odot (y \lor z) = (x \odot y) \lor (x \odot z)$
   \end{tabular}
\end{center}
Notice that these conditions stress the fact that an MV--algebra represents a particular weakening
of a Boolean algebra, where $\oplus$ and $\odot$ are generally non idempotent.
Also in any MV--algebra a partial order relation can be induced making use of \eqref{eq:mv-ord}.
%

\begin{theorem}
The structure $\parang{L, \land,\lor, \lnot, 0, 1}$ is a {\em Kleene lattice\/} (that is a bounded
involutive distributive lattice satisfying the Kleene condition).
\end{theorem}
%

As a general consequence of this result, since in the many--valued case $\oplus$ and $\odot$
together with $\neg$ can express $\lor$ and $\land$, the (additive) operations $\oplus$ and $\odot$
are regarded as more fundamental than the lattice operations.
Every MV--algebra is a subdirect product of totally ordered MV--algebras and an equation holds in the class of all
MV--algebras if and only if it holds in the MV--algebra based on $L_{\aleph_1}$ (\cite{chang-58}).
%
%
Actually the proof of completeness of finite valued logics needs stronger structures; for this
purpose R.~Grigolia (\cite{grigolia}) introduced MV$_d$--algebras which are particular kinds of
MV--algebras.

%
Let us recall that a Wajsberg (W) algebra is a system $\para{A,\to_L,\neg,1}$ where $L$ is a
nonempty set, $1$ is a constant element, $\to_L$ is a binary operation, and $\neg$ is a unary
operation, satisfying conditions $(BZW1)$--$(BZW4)$ of Definition \ref{df:bzw}. The two structures
of Chang and of Wajsberg many--valued algebras are categorically equivalent according to the
following straightforward result.
\begin{theorem}
\par\noindent
\begin{enumerate}
\item
Let ${\bf L}=\para{L,\oplus,\neg,0}$ be a Chang MV--algebra. Then putting $x\to_L y=\neg x\oplus y$
and $1=\neg 0$ the corresponding structure ${\bf L}^W=\para{L,\to_L,\neg,1}$ is a Wajsberg
MV--algebra.
\item
Let ${\bf A}=\para{A,\to_L,\neg,1}$ be a Wajsberg MV--algebra. Then putting $x\oplus y=\neg x\to_L
y$ and $0=\neg 1$ the corresponding structure ${\bf A}^C=\para{A,\oplus,\neg, 0}$ is a Chang
MV--algebra.
\item
Let ${\bf L}$ be a Chang MV--algebra then $\parto{{\bf L}^{W}}^C={\bf L}$ and let ${\bf A}$ be a
Wajsberg MV--algebra then $\parto{{\bf A}^L}^C={\bf A}$.
\end{enumerate}
\end{theorem}
\newpage

\end{document}